\documentclass[12pt]{article}
\usepackage{amsfonts}
\usepackage{amsmath}
\usepackage{amssymb}
\usepackage{graphicx}
\usepackage{color}
\usepackage[all, knot]{xy}
\usepackage{tikz}
\usepackage{braket}

\usepackage[utf8]{inputenc}

\usepackage{epstopdf}
\usepackage[footnotesize]{caption}
\usepackage{amsthm}
\usepackage{enumitem}
\usepackage{mathrsfs}

\usepackage{mathtools}     %% Ò»Ð©ÊýÑ§·ûºÅ
\usepackage{subeqnarray}          %% ÅäºÏcases ¶Ô´óÀ¨ºÅ
\usepackage{cases}                %% ÄÚµÄÊ½×Ó·Ö±ð±àºÅ
\usepackage{color}
\usepackage{subfigure}
\usepackage{cite}                %%ÊÇÒýÓÃ¸ñÊ½±äÆ¯ÁÁ  %%PRDµÈAPS Ä£°å±ØÐë½ûÓÃ¸ÃÏî
\usepackage{hyperref}            %%³¬Á´½Ó±ØÐë
\usepackage{multirow,makecell}   %%±í¸ñ
\usepackage{textcomp}
\usepackage{wasysym}
\usepackage{url}

\usepackage[margin=3cm]{geometry}

\usepackage{geometry}
\usepackage{amsfonts}
\usepackage{graphicx}
\usepackage{amsmath}
\usepackage{hyperref}
\usepackage{tensor}
\usepackage{subfigure}
\usepackage{cases}
\usepackage{appendix}
\usepackage{multirow}
\usepackage{color}
\definecolor{deepblue}{HTML}{003399}
\definecolor{pinkred}{HTML}{ff0066}

\def\[#1\]{\begin{align}#1\end{align}}

\def \nn {\nonumber}

\def \r {\rho}

\def \d {\delta}

\def \dd{\mathrm{d}}

\def \e {\epsilon}

\def \f {\phi}
\def \m {\mu}
\def \n {\nu}
\def \et{\eta}
\def \etb{\bar{\eta}}
\def \pd{\partial}
\def \pdb{\bar{\pd}}

\def \l{\lambda}
\def \s{\sigma}
\def \t{\tau}
\def\ra{\rangle}
\def\la{\langle}

\def \eb{\bar{1}}

\def \mo{\mathcal{O}}

\def \zb{\bar{z}}

\def \intt{\int_{\text{T}^2}\dd^2x }

\def \oa{\int_{z_0}^{z_0+2w}}
\def \ab{\int_{z_0+2w}^{z_0+2w+2w'}}
\def \bc{\int_{z_0+2w+2w'}^{z_0+2w'}}
\def \co{\int_{z_0+2w'}^{z_0}}
\def \pb{\bar{P}}
\def \zetab{\bar{\zeta}}
\def \tf{\text{T}^{2}}
\def \limgr{\lim\limits_{\e\rightarrow0}G(\e)}
\def \limr{\lim\limits_{\e\rightarrow0}}

\def \L{\mathcal{L}}

\def \Ln{\L^{(n)}}
\def \Tn{T_{\m\n}^{(n)}}
\def \gmn{g_{\m\n}}

\def \igmn{g^{\m\n}}

\def \gm{\gamma^\m}

\def \ga{\gamma^a}

\def \hxab{\hat{X}_{ab}}

\def \xbl{X^{b}_{~\l}}
\def \xcl{X^c_{~\l}}
\def \hx{\hat{X}}
\def \dab{\delta_{ab}}

\def \pda{\pd_a}
\def \ema{e^\m_{~a}}

\def \enb{e^\n_{~b}}
\def \ela{e^\l_{~a}}

\def \elc{e^\l_{~c}}

\def \TR{T^{(2)}_{\m\n}}
\def \LO{\L^{(0)}}
\def \LE{\L^{(1)}}
\def \LR{\L^{(2)}}
\def \LTO{T^{(0)}_{ab}}
\def \LTE{T^{(1)}_{ab}}
\def \LTR{T^{(2)}_{ab}}

\def \TR{\text{Tr}}
\def \ppsi{\Psi}
\def \ppsib{\bar{\Psi}}
\def \pse{\psi}
\def \pseb{\psi^*}
\def \psr{\bar{\psi}}
\def \psrb{\bar{\psi}^*}
\def \dpz{\overleftrightarrow{\pd}}
\def \dpzb{\overleftrightarrow{\bar{\pd}}}
\def \pz{\pd}
\def \pzb{\bar{\pd}}

\def \zer{z_{12}}

\def \T{\text{T}^2}
\def \Tbar{\bar{T}}
\def \ete{\et_1}
\def \eteb{\etb_1}
\def \tb{\bar{\t}}
\def \pdt{\pd_{\t}}
\def \pdtb{\pd_{\tb}}
\def \intm{\int_{\mathcal{M}}\sqrt{g}\dd^2 x}
\def \pn{P_{\n}}
\def \zber{\bar{z}_{12}}
\def \pnb{\bar{P}_\n}
\def \en{e_{\n-1}}
\def \enbar{\bar{e}_{\n-1}}
\def \llv{\left\lvert}
\def \rrv{\right\rvert}
\def \inter{\int_{\T_1}\int_{\T_2}}

\def \tor{\text{tor}}
\def \FB{\text{FB}}
\def \DF{\text{DF}}
\def \MF{\text{MF}}
\def \o{\text{o}}
\def \Z{\mathcal{Z}}
\begin{document}
\begin{titlepage}
%\begin{flushright}
%TIT/HEP-6XX \\
%mm,  2016
%\end{flushright}
\vspace{0.5cm}
\begin{center}
{\Large \bf {$T\bar{T}$-flow effects on torus partition functions }}

\lineskip .75em
\vskip 2.5cm
{\large  Song He$^{a,b,}$\footnote{hesong@jlu.edu.cn}, Yuan Sun$^{a,}$\footnote{sunyuan@jlu.edu.cn} ,Yu-Xuan Zhang$^{a,}$\footnote{yuxuanz18@mails.jlu.edu.cn}}
\vskip 2.5em
{\normalsize\it $^{a}$Center for Theoretical Physics and College of Physics, Jilin University,\\ Changchun 130012, People's Republic of China\\
 $^{b}$Max Planck Institute for Gravitational Physics (Albert Einstein Institute),\\
Am M\"uhlenberg 1, 14476 Golm, Germany\\
}
\vskip 3.0em
\end{center}
\begin{abstract}
In this paper, we investigate the partition functions of conformal field theories (CFTs) with the $T\bar{T}$ deformation on a torus in terms of the perturbative QFT approach. In Lagrangian path integral formalism, the first- and second-order deformations to the partition functions of 2D free bosons, free Dirac fermions, and free Majorana fermions on a torus are obtained. The corresponding Lagrangian counterterms in these theories are also discussed. The first two orders of the deformed partition functions and the first-order vacuum expectation value (VEV) of  the first quantum KdV charge obtained by the perturbative QFT approach are consistent with results obtained by the Hamiltonian formalism in literature.
\end{abstract}
\end{titlepage}

\baselineskip=0.7cm

\tableofcontents
\section{Introduction}

The $T\bar{T}$ deformation of field theory has attracted much research interest in recent years both from viewpoint of field theory and in the context of holographic duality. The $T\bar{T}$ deformation of 2D field theory is typically defined on the plane or cylinder by  \cite{Smirnov:2016lqw,Cavaglia:2016oda}
\[
\frac{\dd\L^\l}{\dd\l}=\frac{1}{2}\e^{\m\n}\e^{\r\s} T^\l_{\m\r}T^\l_{\n\s},\label{flow}
\]
where $T^\l$ depending on $\l$ is stress tensor of the theory $\L^\l$. Though the RHS is a composite operator, it is well-defined quantum mechanically \cite{Zamolodchikov:2004ce}.
Remarkably, the $T\bar{T}$ deformation  keeps the integrability of the un-deformed theory and the deformed theory is solvable in some sense \cite{LeFloch:2019wlf,Jorjadze:2020ili,Smirnov:2016lqw,Cavaglia:2016oda,Datta:2018thy,Aharony:2018bad,Rosenhaus:2019utc}.
Since the deformation is irrelevant, the density of states of the deformed theory in the UV exhibits Hagedorn growth behavior, which implies {the $T\bar{T}$ deformation} is non-local in  the UV \cite{Cavaglia:2016oda,Giveon:2017nie,Jiang:2019hxb}. With many intriguing properties discovered, the  $T\bar{T}$ deformation has subsequently been generalized to many directions, {for instance}, to other integrable deformations such as {the $J\bar{T}$ deformation} \cite{Guica:2017lia,Bzowski:2018pcy,Conti:2019dxg}, to supersymmetric cases \cite{Baggio:2018rpv,Chang:2018dge,Jiang:2019trm,Chang:2019kiu}, to various dimensions \cite{Taylor:2018xcy,Gross:2019ach,Gross:2019uxi,Chakraborty:2020xwo} and spin chain models \cite{Bargheer:2008jt,Bargheer:2009xy,Marchetto:2019yyt,Pozsgay:2019ekd,Jiang:2020nnb}. For some other developments of {the $T\bar{T}$ deformation}, please refer to \cite{Mussardo:1999aj,Dubovsky:2017cnj,Cardy:2018sdv,Dubovsky:2018bmo,Jiang:2019tcq,Asrat:2020uib,Santilli:2020qvd,Chakraborty:2020udr,Ouyang:2020rpq}.

Among these progresses, {the partition functions} as well as {correlation functions} in  deformed {CFTs are} of particular interest in our present study. {The partition functions of the   $T\bar{T}$ deformed CFTs} have been computed in \cite{Datta:2018thy} by using the known deformed spectrum. {Since the results in \cite{Datta:2018thy} are nonperturbative, the modular properties can be discussed, and it was shown that the partition functions are modular covariant.} From other perspective,  {the deformed partitions were discussed from random metric point of view \cite{Cardy:2018sdv}, and also  in the context of holographic duality \cite{Caputa:2019pam}.} As for {correlation functions}, {the deformed one-point functions of KdV charges operators were  considered non-perturbatively based on the deformed spectrum} \cite{Asrat:2020jsh}. Also the general deformed correlation functions in the UV {were} considered by J. Cardy in \cite{Cardy:2019qao}.

On the other hand, one can study the $T\bar{T}$ deformation in a perturbative way. More concretely, {suppose that one can expand $\L^\l$ around $\l=0$,}
\[
\L^\l=\L^{(0)}+\l \L^{(1)}+\frac{\l^2}{2!}\L^{(2)}+...,
\]
where the first term $\L^{(0)}$ corresponds to the un-deformed theory, the second term is the $T\bar{T}$ operator of un-deformed theory as appeared in the RHS of (\ref{flow}) with $\l=0$, the third term and the terms omitted are presented since the stress tensor $T^\l$ is not fixed but also flow under the deformation. In other words, the stress tensor depends on $\l$.

A number of works were done in the framework of perturbation method, for example, in \cite{Smirnov:2016lqw} the renormalization of free theory under { the $T\bar{T}$ deformation} is investigated by matching the S-matrix. Meanwhile, other physical quantities were also computed perturbatively, such as entanglement entropies, wilson loop and correlation functions \cite{Jeong:2019ylz,Sun:2019ijq,Chakraborty:2018aji}. In this work, we will continue to study the {partition functions} (which can be treated as zero-point {functions}) of deformed {CFTs} in a perturbative manner. The {correlation functions} of deformed {theories were} considered earlier in \cite{Guica:2019vnb,Kraus:2018xrn, Aharony:2018vux}, where {two-point  functions and three-point functions} were calculated, as well as the correlation functions of stress tensors. Later, these results were generalized to higher-point function cases \cite{He:2019vzf,He:2020qcs}, as well as including supersymmetry \cite{He:2019ahx}, torus {CFTs} \cite{He:2020udl}, and especially the holographic dual of stress tensor correlation function in large $c$ limit {was} considered in  \cite{Li:2020pwa}.

In these studies of correlation functions, it is worthwhile to note that computation is mainly performed in the first-order perturbation of CFT or in the case where the CFT is defined on the plane. Naturally, to make progress, the next step is that can we go beyond the first-order perturbation.  However, this is a nontrivial question as can be seen as follow. As discussed above, in the first-order perturbation, the $T\bar{T}$ operator is known which is just constructed from the stress tensor of the un-deformed CFT, while in higher-order perturbations, one must take the corrections of $T\bar{T}$ operator into consideration, namely, $T\bar{T}$-flow effects. Unfortunately, in a general CFT, we do not have such an explicit expression on such kinds of corrections. Nevertheless, as the first step towards higher-order perturbations, we can start with free theory, where the corrections of stress tensor and Lagrangian under { the $T\bar{T}$ deformation} can be constructed explicitly order by order. Based on this setup, we will study the {corrections of deformed partition functions} up to second-order by employing {perturbation method}. This also {generalize} our previous work \cite{He:2020udl}, where the first-order partition {functions of deformed CFTs on torus were computed}. Moreover, since we work in free theories,  we will use Wick contraction rather than the Ward identity obtained in \cite{He:2020udl} to figure out the deformed correlation functions. Finally, the two methods will lead to the same results.

The organization of this paper is as follows. In Section \ref{sec2}, we review the general method to {obtain} the deformed Lagrangian and stress tensor order by order, which can be used to expand the partition function up to the second-order that we are interested in. In Section \ref{secBS}, Section \ref{secDF} and Section \ref{secMF}, we computed the first- and second-order corrections to the partition functions of free bosons, Dirac fermions and Majorana fermions respectively. We use Wick contraction to computed the deformed partition functions, also some proper regularization methods are chosen. In Section \ref{secKDV} we continue to calculate the VEV of the first KdV charge in the deformed free theories up to the first-order, by using the perturbative QFT approach. We end in Section \ref{smy} with a conclusion and discussion.  Our conventions, useful formulae, and some calculation details are presented in the appendices.
%\section{Preliminary}
\section{$T\Tbar$ deformed partition function for generic 2d theory}\label{sec2}
In this section, we would like to compute the perturbation expansion of $T\bar{T}$ deformed partition function beyond the first-order. The procedure is based on the method first introduced in \cite{Cavaglia:2016oda} (also see \cite{Bonelli:2018kik}), where deformed Lagrangian is obtained order by order. Let us first review this method below.

Consider a $T\Tbar$ deformed QFT living in a two-dimensional Euclidean spacetime $(\mathcal{M},g_{ab})$ whose dynamics is governed by the local action
\[
S^\l=\intm\mathcal{L}^\l(\phi,\nabla_{a}\phi,g_{ab}).
\]
Here $\L^\l$ denotes the deformed Lagrangian parameterized by $\l$. The $T\bar{T}$ deformation can then be defined by the following flow equation
\[
\frac{\dd\L^\l}{\dd\l}=\frac{1}{2}\e^{\m\n}\e^{\r\s} T^\l_{\m\r}T^\l_{\n\s},\label{flow}
\]
where $\e_{\m\n}=g_{\m\r}g_{\n\s}\e^{\r\s}$  is the volume element of the spacetime, and $T^\lambda_{\m\n}$ is the stress tensor of the deformed theory, which is defined as
\[
T^\lambda_{\m\n}=\frac{2}{\sqrt{g}}\frac{\d S^\l}{\d g^{\m\n}}=2\frac{\pd\L^{\l}}{\pd g^{\m\n}}-g_{\m\n}\L^\l.\label{stress}
\]
%Since our aim in this subsection is to obtain the analytic form of $\L^\l$ by solving Eq.\eqref{flow}, we next review the approach presented in \cite{Cavaglia:2016oda} to realize it. Firstly  we set
Now expand of deformed Lagrangian and stress tensor in the power of $\l$
\[
\L^\l=\sum_{n=0}^{\infty}\frac{\l^n}{n!}\Ln,~~~~T^\l_{\m\n}=\sum_{n=0}^{\infty}\frac{\l^n}{n!}\Tn.\label{expand}
\]
In order to figure out $\L^{(n)}$, one can plugging \eqref{expand} into both \eqref{flow} and \eqref{stress}. By comparing each order in the resulting expressions, eventually, we obtain the following recursion relations \footnote{The identity $g^{\m\n}g^{\r\s}-g^{\r\n}g^{\m\s}=\e^{\m\r}\e^{\n\s}$ is used.}
\[
\L^{(n+1)}=&\frac{1}{2}\sum_{i=0}^{n}C^{i}_{n}\Big(T^{\m(i)}_{~~\m}T^{\n(n-i)}_{~~\n}-T^{\m(i)}_{~~\n}T_{~~\mu}^{\n(n-i)}\Big),\label{Ln}\\
\Tn=&2\frac{\pd\Ln}{\pd g^{\m\n}}-g_{\m\n}\Ln,\label{Tn}
\]
where $C^{i}_{n}\equiv\frac{n!}{i!(n-i)!}$. Note this recursion relations allow us to obtain $\L^{(n)}$ and $T^{(n)}_{\m\n}$ for arbitrary $n$, once $\L^{(0)}$, i.e. the un-deformed theory, is given.
%\subsection{Deformed partition function}

With perturbations of $\L^{\l}$ acquired, we continue to derive the corrections of the partition function to higher-order in perturbation theory in path integral language, which is
\[
\Z^{\l}&=\int\mathcal{D}\phi~\text{e}^{-\int_{\mathcal{M}}\L^{\l}[\phi]}\nn\\
&=\Z^{(0)}-\l\Z^{(0)}\int_{\mathcal{M}}\la \L^{(1)}\ra+\frac{\l^2}{2}\Z^{(0)}\big(\int_{\mathcal{M}}\int_{\mathcal{M}}\la\L^{(1)}(x)\L^{(1)}(x')\ra-\int_{\mathcal{M}}\la\L^{(2)}\ra\big)+\mo(\l^3)\nn\\
&\equiv \Z^{(0)}+\l \Z^{(1)}+\frac{\l^2}{2}\Z^{(2)}+O(\l^3),
\]
where
\[
\Z^{(0)}=&\int\mathcal{D}\phi~\text{e}^{-\int_{\mathcal{M}}\L^{(0)}[\f]},\\
\Z^{(1)}=&-\Z^{(0)}\int_{\mathcal{M}}\la \L^{(1)}\ra,\label{zz1}\\
\Z^{(2)}=&\Z^{(0)}\big(\int_{\mathcal{M}}\int_{\mathcal{M}}\la\L^{(1)}(x)\L^{(1)}(x')\ra-\int_{\mathcal{M}}\la\L^{(2)}\ra\big).\label{zz2}
\]
In what follows, we will focus on the $T\bar{T}$ deformed free theories on a torus, including free bosons, Dirac fermions, and Majorana fermions, where the deformed partition functions up to the second-order (\ref{zz1}--\ref{zz2}) can be worked out analytically.
\section{Free bosons}\label{secBS}
At first, what we would like to consider is the $T\bar{T}$ deformed free scalar on a torus $\text{T}^2$. The corresponding action of the un-deformed theory reads
\[
S=\frac{g}{2}\intt\pd_\m\phi\pd^\m\phi,
\]
where $g$ is a normalization constant.
According to the recursion relations (\ref{Ln}-\ref{Tn}) mentioned above, one could obtain the deformed Lagrangian and stress tensor starting from $\L^{(0)}$,
\[
\LO&=2g\pd\phi\pdb\phi.
\]
Then the un-deformed stress tensor is\footnote{In this paper, we use the conventional notation that $T\equiv -2\pi T_{zz},~\bar{T}\equiv -2\pi T_{\zb\zb}$, and $\Theta\equiv 2\pi T_{z\zb}$. The complex coordinates $z:=x+iy$, where $y$ is Euclidean time. $\partial:=(\partial_{x}-i\partial_{y})/2$. The metric $g_{z\bar{z}}=\frac12$. }
\[
T^{(0)}&=-2\pi g(\pd\f)^2,\quad\bar{T}^{(0)}=-2\pi g(\pdb\f)^2,\quad\Theta^{(0)}=0,
\]
from which the first-order Lagrangian is given by
\[
\LE=-\frac{1}{\pi^2}T^{(0)}\bar{T}^{(0)}=-4g^2(\pd\phi\pdb\phi)^2,\label{bsh1}
\]
and the corresponding first-order  stress tensor  is
\[
T^{(1)}&=8\pi g^2(\pd\f)^3(\pdb\f),\quad \bar{T}^{(1)}=8\pi g^2(\pdb\f)^3(\pd\f),\quad\Theta^{(1)}=-4\pi g^2(\pd\f\pdb\f)^2.\label{deformedboson}
\]
Reusing Eq.(\ref{Ln}), we end up with the second-order Lagrangian
\[
\LR=-\frac{1}{\pi^2}(T^{(0)}\bar{T}^{(1)}+\bar{T}^{(0)}T^{(1)})=32g^3(\pd\f\pdb\f)^3,\label{bsh2}
\]
 We then could write out the corrections of the partition function \eqref{zz1} and \eqref{zz2} more concretely for bosonic fields
\[
\Z^{(1)}=&\frac{1}{\pi^2}\Z^{(0)}\int_{\T}\la T\Tbar^{(0)}(z,\zb)\ra=4g^2\Z^{(0)}\int_{\T}\la(\pd\f\pdb\f)^2\ra\label{z1},\\
\Z^{(2)}=&\frac{1}{\pi^4}\Z^{(0)}\int_{\T_1}\int_{\text{T}_2^2}\la T\Tbar^{(0)}(z_1,\zb_1) T\Tbar^{(0)}(z_2,\zb_2)\ra+\frac{1}{\pi^2}\Z^{(0)}\int_{\T}\la T^{(0)}\Tbar^{(1)}+T^{(1)}\Tbar^{(0)}\ra\nn\\
=&16g^4\Z^{(0)}\int_{\T_1}\int_{\text{T}_2^2}\la(\pd_1\f\pdb_1\f)^2(\pd_2\f\pdb_2\f)^2\ra-32g^3\Z^{(0)}\int_{\T}\la(\pd\f\pdb\f)^3\ra\label{z2}.
\]
Note that the expectation values in (\ref{z1}-\ref{z2}) are defined in free theory,  all of them could be evaluated directly by applying Wick contraction since the propagator is well-known for torus free scalar field\cite{DiFrancesco:1997nk},
\[
\la\f(z_1,\zb_1)\f(z_2,\zb_2)\ra=&(4\pi g)^{-1}\Big(-\log\left\lvert\frac{\vartheta_1(\zer)}{\et(\t)}\right\rvert^2+2\pi\frac{(\text{Im}[\zer])^2}{\t_2}\Big).\label{boson propagatos}
\]
Here {$\vartheta_1(z)$  is one of Jacobi $\vartheta$-functions and $\eta(\tau)$ is Dedekind $\eta$-function}. Performing derivatives on \eqref{boson propagatos} gives {various two-point functions}\footnote{We neglect the last term of Eq.\eqref{deltafunc} in the later part of this paper, since $\d^{(2)}\big(z_{12}-(m+n\t)\big)$ is always zero in the integral region we're considering. We did the same thing later on when we deal with fermionic fields.}
\[
\la\pd\f(z_1,\zb_1)\pd\f(z_2,\zb_2)\ra=&(4\pi g)^{-1}\big(\frac{\pi}{\t_2}-2\eta_1-P(\zer)\big),\\
\la\pdb\f(z_1,\zb_1)\pdb\f(z_2,\zb_2)\ra=&(4\pi g)^{-1}\big(\frac{\pi}{\t_2}-2\bar{\eta}_1-\bar{P}(\zb_{12})\big),\\
\la\pd\f(z_1,\zb_1)\pdb\f(z_2,\zb_2)\ra=&(4\pi g)^{-1}\Big(\pi\d^{(2)} (z_{12})-\frac{\pi}{\t_2}+\sum_{\substack{\{m,n\}\neq\{0,0\}}}\pi\d^{(2)}\big(z_{12}-(m+n\t)\big)\Big),\label{deltafunc}
\]
where $ P(z)$ is Weierstrass  elliptic function and we have applied the formula $\pzb (z^{-1})=\pz(\zb^{-1})=\pi\d^{(2)}(\vec{x})\equiv\pi\d^{(2)}(z)$. For  more details on elliptic functions please refer to  Appendix \ref{a}.  The subsequent derivation of Wick contraction indicates that  the expectation values of the composite operators  $\big(\pd\f(z_1,\zb_1)\big)^2$, $\big(\pdb\f(z_1,\zb_1)\big)^2$, and $\left\lvert\pd\f(z_1,\zb_1)\right\rvert^2$ also make contributions. We regularize them by utilizing the point-splitting method
\[
\la\pd\f(z_1,\zb_1)\pd\f(z_1,\zb_1)\ra&=\lim\limits_{z_2\rightarrow z_1}\Big(\la\pd\f(z_1,\zb_1)\pd\f(z_2,\zb_2)\ra+\frac{1}{4\pi g\zer^2}\Big)=(4\pi g)^{-1}\big(\frac{\pi}{\t_2}-2\eta_1\big),\\
\la\pdb\f(z_1,\zb_1)\pdb\f(z_1,\zb_1)\ra&=\lim\limits_{z_2\rightarrow z_1}\Big(\la\pdb\f(z_1,\zb_1)\pdb\f(z_2,\zb_2)\ra+\frac{1}{4\pi g\zb_{12}^2}\Big)=(4\pi g)^{-1}\big(\frac{\pi}{\t_2}-2\bar{\eta}_1\big),\\
\la\pd\f(z_1,\zb_1)\pdb\f(z_1,\zb_1)\ra&=\lim\limits_{z_2\rightarrow z_1}\Big(\la\pd\f(z_1,\zb_1)\pdb\f(z_2,\zb_2)\ra-\frac{\d(\zer)}{4 g}\Big)=\frac{-1}{4g\t_2}.
\]
With all ingredients in place, we next go on to investigate the corrections to the partition function of free bosons.
\subsection{First-order}
First, we note that the partition function of the free scalar on a torus is
\[
\Z^{(0)}=\frac{1}{\sqrt{\t_2}|\et(\t)|^2}.
\]
 According to Eq.(\ref{z1}), at the first-order we shall just compute the  value of $\intt\la T\Tbar^{(0)}(z,\zb)\ra$, \footnote{Here $i\equiv\pd\f(z_i,\zb_i)$, $\bar{i}\equiv\pdb\f(z_i,\zb_i)$, $(i=1,2,3...).$}
\[
\int_{\T_1}\dd^2x_1 \la T\Tbar^{(0)}(z_1,\zb_1)\ra=&4\pi^2g^2\t_2\la\pd\f(z_1,\zb_1)\pd\f(z_1,\zb_1)\pdb\f(z_1,\zb_1)\pdb\f(z_1,\zb_1)\ra\nn\\
=&4\pi^2g^2\t_2\big(2\la1\bar{1}\ra^2+\la11\ra\la\eb\eb\ra\big)\nn\\
=&\frac{3\pi^2}{4\t_2}+\lvert\eta_1\rvert^2\t_2-\frac{\pi}{2}(\ete+\eteb)\nn\\
=&\frac{4\pi^2}{\Z^{(0)}}\t_2\pdt\pdtb \Z^{(0)},\label{bonson1}
\]
which is consistent with \cite{He:2020udl}. Thus the first-order correction of the partition function is
\[
\Z^{(1)}=\frac{\Z^{(0)}}{\pi^2}\int_{\T_1}\dd^2x_1 \la T\Tbar^{(0)}(z_1,\zb_1)\ra=4\t_2\pdt\pdtb \Z^{(0)}\label{boson1ordr}.
\]
\subsection{Second-order}
We next go on to consider the second-order correction to the partition function. We begin with calculating the first term of (\ref{z2}), whose integrand can be contracted as\footnote{ To obtain Eq.\ref{tttt}, we have discarded terms purely divergent(i.e., they have no finite contribution to the final result under the minimal subtraction principle). }
\[
&\la T\Tbar^{(0)}(z_1,\zb_1) T\Tbar^{(0)}(z_2,\zb_2)\ra=(2\pi g)^4\la11\bar{1}\bar{1}22\bar{2}\bar{2}\ra\nn\\
=&(2\pi g)^4\Big[\la11\ra\la\bar{1}\bar{1}\ra\la22\ra\la\bar{2}\bar{2}\ra+2\times\big(\la11\ra\la\bar{1}\bar{1}\ra\la2\bar{2}\ra^2+\la11\ra\la22\ra\la\bar{1}\bar{2}\ra^2+\la11\ra\la\bar{2}\bar{2}\ra\la\bar{1}2\ra^2+\la\bar{1}\bar{1}\ra\la22\ra\la1\bar{2}\ra^2\nn\\
&+\la\bar{1}\bar{1}\ra\la\bar{2}\bar{2}\ra\la12\ra^2+\la22\ra\la\bar{2}\bar{2}\ra\la1\bar{1}\ra^2\big)+8\times\big(\la11\ra\la\bar{1}2\ra\la2\bar{2}\ra\la\bar{2}\bar{1}\ra+\la\bar{1}\bar{1}\ra\la12\ra\la2\bar{2}\ra\la\bar{2}1\ra\nn\\
&+\la22\ra\la1\bar{1}\ra\la\bar{1}\bar{2}\ra\la\bar{2}1\ra+\la\bar{2}\bar{2}\ra\la1\bar{1}\ra\la\bar{1}2\ra\la21\ra\big)+4\times\big(\la1\bar{1}\ra^2\la2\bar{2}\ra^2+\la12\ra^2\la\bar{1}\bar{2}\ra^2+\la\bar{1}2\ra^2\la1\bar{2}\ra^2\big)\nn\\
&+16\times\big(\la1\bar{1}\ra\la\bar{1}2\ra\la2\bar{2}\ra\la\bar{2}1\ra+\la1\bar{1}\ra\la\bar{1}\bar{2}\ra\la\bar{2}2\ra\la21\ra+\la1\bar{2}\ra\la\bar{2}\bar{1}\ra\la\bar{1}2\ra\la21\ra\big)\Big]\nn\\
=&\frac{1}{16}\Big(24A^4+8A^2\lvert B\lvert^2+\lvert B\lvert^4+4\lvert B-P(z_{12})\lvert^4+32A^2\lvert B-P(z_{12})\lvert^2+4\cdot\text{Re}\big[\bar{B}^2(B-P(z_{12}))^2\big]\nn\\
&+32A^2\cdot\text{Re}\left[\bar{B}(B-P(z_{12}))\right]-48\pi A^3\delta(z_{12})-72\pi A\lvert B\lvert^2\delta(z_{12})\Big),\label{tttt}
\]
where $B\equiv(\frac{\pi}{\t_2}-2\eta_1)$, $\bar{B}\equiv(\frac{\pi}{\t_2}-2\eteb)$, and $A\equiv\frac{\pi}{\t_2}$.
Integrating the above expression  amounts to compute the following integrals
\[
\int_{\T_1}\int_{\T_2}\big(B-P(z_{12})\big)&=0,\label{pms}\\
\int_{\T_1}\int_{\T_2}\big(B-P(z_{12})\big)^2&=\frac{g_2\t_2^2}{12}-\t_2^2B^2,\\
\int_{\T_1}\int_{\T_2}\llv (B-P(z_{12})\rrv^2&=-\pi^2,\\
\int_{\T_1}\int_{\T_2}\llv B-P(z_{12})\rrv^4&=\t_2^2|B|^4+\frac{|g_2|^2\t_2^2}{12^2}-4\t_2^2A^2|B|^2-B^2\frac{\bar{g}_2\t_2^2}{12}-\bar{B}^2\frac{g_2\t_2^2}{12}\label{ps},
\]
where $g_2$ is one of  Weierstrass invariants {whose definition can be found in Appendix \ref{a}. We collect the detailed computation of the above integrals in Appendix \ref{b.1}. Note some of the integrals are divergent, thus a proper regularization scheme is needed, which will be presented in Appendix \ref{b.0}.

With the help of (\ref{tttt}$-$\ref{ps}) and the following identity  relating the quantity $g_2$ with $\ete$
\[
g_2=48\big(i\pi\pdt\eta_1+\ete^2\big),\label{gone}
\]
the double integral of Eq.\eqref{tttt} is derived as
\[
&\int_{\T_1}\int_{\text{T}_2^2}\la T\Tbar^{(0)}(z_1,\zb_1) T\Tbar^{(0)}(z_2,\zb_2)\ra\nn\\
=&\t_2^2|\ete|^4-\pi^2|\ete|^2+4\pi^2\t_2^2|\pdt\ete|^2-\frac{\pi^2}{4}(\ete^2+\eteb^2)+\frac{3\pi^3}{4\t_2}(\ete+\eteb)+\t_2\pi|\ete|^2(\ete+\eteb)\nn\\
&+\frac{i\pi^3}{2}(\pdtb\eteb-\pdt\ete)+2i\pi^2\t_2(\eteb\pdt\ete-\ete\pdtb\eteb)+2i\pi\t_2^2(\eteb^2\pdt\ete-\ete^2\pdtb\eteb)-\frac{15\pi^4}{2^4\t_2^2}\nn\\
&-18\pi^2|\ete|^2+\frac{9\pi^3}{\t_2}(\ete+\eteb)-\frac{15\pi^4}{2\t_2^2}\nn\\
=&\frac{16\pi^4}{\Z^{(0)}}\big(\t_2^2\pdt^2\pdtb^2+i\t_2(\pdt^2\pdtb-\pdtb^2\pdt)\big)\Z^{(0)}-\frac{72\pi^4}{\Z^{(0)}}\pdt\pdtb \Z^{(0)}+\frac{6\pi^4}{\t_2^2}.\label{3927}
\]
Consequently,
\[
&\Z^{(0)}\int_{\T_1}\int_{\text{T}_2^2}\la \L^{(1)}(z_1,\zb_1)\L^{(1)}(z_2,\zb_2)\ra\nn\\
=&\frac{1}{\pi^4}\Z^{(0)}\int_{\T_1}\int_{\text{T}_2^2}\la T\Tbar^{(0)}(z_1,\zb_1) T\Tbar^{(0)}(z_2,\zb_2)\ra\nn\\
=&16\big(\t_2^2\pdt^2\pdtb^2+i\t_2(\pdt^2\pdtb-\pdtb^2\pdt)\big)\Z^{(0)}-72\pdt\pdtb \Z^{(0)}+6\t_2^{-2}\Z^{(0)}.\label{24}
\]
%which is in good agreement with the result in \cite{He:2020udl}(version 1).

We next move to evaluate the second term in \eqref{z2}. Using Wick contraction, the integrand is
\[
&\la T^{(0)}(z_1,\zb_1)\Tbar^{(1)}(z_1,\zb_1)\ra+\la T^{(1)}(z_1,\zb_1)\bar{T}^{(0)}(z_1,\zb_1)\ra\nn\\
=&-32\pi^2g^3\la111\bar{1}\bar{1}\bar{1}\ra\nn\\
=&-288\pi^2g^3\times\la1\bar{1}\ra\la11\ra\la\bar{1}\bar{1}\ra-192\pi^2g^3\times\la1\bar{1}\ra^3\nn\\
=&\frac{18}{\t_2}|\ete|^2-\frac{9\pi}{\t_2^2}(\ete+\eteb)+\frac{15\pi^2}{2\t_2^3}\nn\\
=&\frac{72\pi^2}{\t_2\Z^{(0)}}\pdt\pdtb \Z^{(0)}-\frac{6\pi^2}{\t_2^3}.
\]
After simple integration, one has
\[
\Z^{(0)}\int_{\text{T}^2}\L^{(2)}(z,\zb)=
-\frac{1}{\pi^2}\Z^{(0)}\int_{\T}\la T^{(0)}\Tbar^{(1)}+T^{(1)}\Tbar^{(0)}\ra=-72\pdt\pdtb \Z^{(0)} +6\t_2^{-2}\Z^{(0)}\label{26}.
\]
Putting together \eqref{24} and \eqref{26}, we obtain the second-order correction of the partition function under the $T\bar{T}$ deformation
\[
\Z^{(2)}=&\Z^{(0)}\int_{\T_1}\int_{\text{T}_2^2}\la \L^{(1)}(z_1,\zb_1)\L^{(1)}(z_2,\zb_2)\ra-\Z^{(0)}\int_{\text{T}^2}\L^{(2)}(z,\zb)\nn\\
=&16\big(\t_2^2\pdt^2\pdtb^2+i\t_2(\pdt^2\pdtb-\pdtb^2\pdt)\big)\Z^{(0)}\label{boson2order},
\]
which is consistent with \cite{Datta:2018thy}.
Note that we have minimally subtracted the divergent terms\footnote{The details of minimal subtraction (\eqref{pzmode},\eqref{pzmodesquare}) are presented in Appendix \ref{b.1} . }  when deriving the RHS of (\ref{3927}), and thus (\ref{boson2order}). It is possible to implement this minimal subtraction  by  adding  the following counterterm \footnote{Please refer to Appendix \ref{sec-ct} for a discussion of this Lagrangian counterterm.}
\[
\L_{\text{FB},\text{ct}}=\l^2\cdot\left\{\frac{8g^2}{\pi\e^2}\big(\pd\phi\pdb\phi\big)^2+\frac{1}{24\pi^3\e^6}\right\}\label{FBct},
\]
where $\e$ stands for the radius of the infinitesimal disk regulator.

\section{Free Dirac fermions}\label{secDF}
For the rest of the examples, we turn our attention to the fermionic fields defined on a torus.
 We first focus on a massless Dirac field whose action is
\[
S=\frac{g}{2}\int_{\T}(\Psi^{\dagger}\gamma^0\gamma^a\pd_a\Psi-\pd_a\Psi^\dagger\gamma^0\gamma^a\Psi)
\]
with
\[
\Psi=[\psi~~\bar{\psi}]^T,\quad\Psi^{\dagger}=[\psi^*~~\bar{\psi}^*].
\]
Our convention for  gamma matrices  are $\{\gamma^{0},\gamma^1\}=\{\sigma^1,\sigma^2\}$, where $\s^i,i=1,2$ are  Pauli matrices.

As before we  make the expansion
\[
\L^\l=&\L^{(0)}+\l \L^{(1)}+...~,\quad T_{\m\n}^\l=T_{\m\n}^{(0)}+\l T_{\m\n}^{(1)}+...~,\label{Ldirac}
\]
remarkably, the case of fermions will simplify a lot comparing with bosons by the fact that the higher-order terms of $\L^{(n)},n\geq2$ are completely vanishing\cite{Bonelli:2018kik}, due to the Grassmannian nature of fermionic fields. Following the derivation presented in \cite{Bonelli:2018kik}, we  obtain the full expression of $\L^\l$ and $T_{\m\n}^\l$ written in complex coordinates \footnote{For the derivation, one can refer to Appendix \ref{c}.}
\[
\L^{(0)}=&g\big(\psi^*\dpzb\psi +\bar{\psi}^*\dpz\bar{\psi}\big),\label{l0dirac}\\
\L^{(1)}=&\frac{1}{\pi^2}\Big(\left(\Theta^{(0)}\right)^2-T^{(0)}\Tbar^{(0)}\Big)\nn\\
=&\frac{g^2}{2}\Big(\big(\psi^*\dpzb\psi\big)\big(\bar{\psi}^*\dpz\bar{\psi}\big)+\big(\psi^*\psi\pzb\psi^*\pzb\psi+\bar{\psi}^*\bar{\psi}\pz\bar{\psi}^*\pz\bar{\psi}\big)\Big)-g^2(\psi^*\dpz\psi)(\bar{\psi}^*\dpzb\bar{\psi}),\label{DiracL1}\\
\L^\l=&\L^{(0)}+\l\cdot\L^{(1)},\label{truncation}
\]
and
\[
T^{(0)}=&-\pi g\cdot\psi^*\dpz\psi,\quad\bar{T}^{(0)}=-\pi g\cdot\bar{\psi}^*\dpzb\bar{\psi},\quad\Theta^{(0)}=-\frac{\pi g}{2}\big(\psi^*\dpzb\psi+\bar{\psi}^*\dpz\bar{\psi}\big),\\
T^{(1)}=&-\frac{\pi g^2}{2}\Big(\psi^*\psi\big(\pzb\psi^*\pz\psi+\pz\psi^*\pzb\psi\big)-\big(\psi^*\dpz\psi\big)\big(\bar{\psi}^*\dpz\bar{\psi}\big)\Big),\\
\bar{T}^{(1)}=&-\frac{\pi g^2}{2}\Big(\bar{\psi}^*\bar{\psi}\big(\pz\bar{\psi}^*\pzb\bar{\psi}+\pzb\bar{\psi}^*\pz\bar{\psi}\big)-\big(\bar{\psi}^*\dpzb\bar{\psi}\big)\big(\psi^*\dpzb\psi\big)\Big),\quad \Theta^{(1)}=0,\label{thetadirac}\\
T^\l_{\m\n}=&T^{(0)}_{\m\n}+\l\cdot T^{(1)}_{\m\n}.
\]

It is well-known that the un-deformed partition function for Dirac fermions is given by
\[
\Z^{(0)}_\n=(d_\n\bar{d}_\n)^2,\quad d_\n(\t)=\left(\frac{{\vartheta_\nu(\t)}}{\eta(\t)}\right)^{1/2}.
\]
where $\nu=1,2,3,4$ denotes the spin structures of fermions, corresponding to different boundary conditions\footnote{$Z^{(0)}_1$ that corresponding to fermions with the double periodic boundary condition is zero, due to the property of Grassmann number \cite{DiFrancesco:1997nk}.}, $\vartheta_\nu$ are Jacobi $\vartheta$-functions. The non-vanishing two-point functions for Dirac fermions with spin structure $\n$ are
 %\footnotetext{Here the function $P_\n(z)$ is defined by \cite{book:1157405}.}
\[
\la\psi^*(z_1)\psi(z_2)\ra_\n=&(2\pi g)^{-1}P_\n(z_{12}),\\
\la\bar{\psi}^*(\zb_1)\bar{\psi}(\zb_2)\ra_\n=&(2\pi g)^{-1}\bar{P}_\n(\zb_{12}),\quad\n=2,3,4.
\]
where
\[
P_\n(z):=\sqrt{P(z)-e_{\n-1}}=\frac{\vartheta_\n(z)\pd_z\vartheta_1(0)}{\vartheta_\n(0)\vartheta_1(z)}.
\]
Performing derivatives on the propagators leads to the following correlation functions
\[
\la\pd\psi^*(z_1)\psi(z_2)\ra_\n=&(2\pi g)^{-1}\pd P_\n(z_{12}),\\
\la\pd\psi^*(z_1)\pd\psi(z_2)\ra_\n=&-(2\pi g)^{-1}\pd^2P_\n(z_{12}),\\
\la\pzb\psi^*(z_1)\psi(z_2)\ra_\n=&(2 g)^{-1}\d^{(2)}(\zer),\\
\la\psi^*(z_1)\pzb\psi(z_2)\ra_\n=&-(2 g)^{-1}\d^{(2)}(\zer).
\]
We need further to regularize these correlation functions when two points coincide with each other, in parallel with the bosonic case, we use the point-splitting method
\[
\la\psi^*(z_1)\psi(z_1)\ra_\n\equiv& \lim\limits_{z_2\rightarrow z_1}\big(\la\psi^*(z_1)\psi(z_2)\ra_\n-(2\pi g z_{12})^{-1}\big)=0,\\
\la\pd\psi^*(z_1)\psi(z_1)\ra_\n\equiv&\lim\limits_{z_2\rightarrow z_1}\big(\la\pd\psi^*(z_1)\psi(z_2)\ra_\n+(2\pi g z_{12}^2)^{-1}\big)=-(4\pi g)^{-1}e_{\n-1},\\
\la\pd\psi^*(z_1)\pd\psi(z_1)\ra_\n\equiv&\lim\limits_{z_2\rightarrow z_1}\big(\la\pd\psi^*(z_1)\pd\psi(z_2)\ra_\n+(\pi gz_{12}^3)^{-1}\big)=0,\\
\la\pzb\psi^*(z_1)\psi(z_1)\ra_\n\equiv&\lim\limits_{z_2\rightarrow z_1}\big(\la\pzb\psi^*(z_1)\psi(z_2)\ra_\n-(2 g)^{-1}\d^{(2)}(z_{12})\big)=0,\\
\la\pzb\psi^*(z_1)\pd\psi(z_1)\ra_\n\equiv&\lim\limits_{z_2\rightarrow z_1}\big(\la\pzb\psi^*(z_1)\pz\psi(z_2)\ra_\n+(2 g)^{-1}\pz \d^{(2)}(z_{12})\big)=0,\\
\la\pzb\psi^*(z_1)\pzb\psi(z_1)\ra_\n\equiv&\lim\limits_{z_2\rightarrow z_1}\big(\la\pzb\psi^*(z_1)\pzb\psi(z_2)\ra_\n+(2g)^{-1}\pzb \d^{(2)}(z_{12})\big)=0.
\]
Now we have all the required ingredients to calculate the corrections to the partition function.

\subsection{First-order}

Using  Wick contraction and the propagators and their derivatives listed above, we can compute the expectation value of  $T^{(0)}\Tbar^{(0)}$ and $(\Theta^{(0)})^2$
\[
\la T^{(0)}\Tbar^{(0)}\ra_\n=\frac{1}{4}\lvert e_{\n-1}\rvert^2=\frac{4\pi^2}{\Z^{(0)}_\n}\pdt\pdtb \Z^{(0)}_\n,\quad\la(\Theta^{(0)})^2\ra_\n=0.
\]
Therefore the first-order correction of the partition function is
\[
\Z^{(1)}_\n=-\int_{\text{T}^2}\L^{(1)}(z,\zb)=\frac{1}{\pi^2}\Z^{(0)}\int_{\T}\la T^{(0)}\Tbar^{(0)}\ra_\n=4\t_2\pdt\pdtb \Z_{\n}^{(0)}.\label{Dira1order}
\]
Note that the first-order correction of free Dirac fermions shares the same structure with that of free bosons \eqref{boson1ordr}, which matches the conclusion in \cite{Datta:2018thy} obtained by the operator formalism. We're going to show that this is also true for the second-order correction.
\subsection{Second-order}
We now proceed to compute the second-order correction. Since there are no higher-order terms in Lagrangian ($\L^{(n)}=0$ for $n\geq2$) for free massless Dirac fermions, (\ref{zz2}) reduces to
\[
\Z^{(2)}_\n=&\Z^{(0)}_\n\int_{\T_1}\int_{\text{T}_2^2}\la \L^{(1)}(z_1,\zb_1)\L^{(1)}(z_2,\zb_2)\ra_{\n}\nn\\
=&\frac{1}{\pi^4}\Z_\n^{(0)}\int_{\T_1}\int_{\T_2}\la\left(\Theta^{(0)}\right)^2(z_1,\zb_1)\left(\Theta^{(0)}\right)^2(z_2,\zb_2)\ra_\n\nn\\
-&\frac{2}{\pi^4}\Z_\n^{(0)}\int_{\T_1}\int_{\T_2}\la T\Tbar^{(0)}(z_1,\zb_1)\left(\Theta^{(0)}\right)^2(z_2,\zb_2)\ra_\n+\frac{1}{\pi^4}\Z_\n^{(0)}\int_{\T_1}\int_{\T_2}\la T\Tbar^{(0)}(x_1) T\Tbar^{(0)}(x_2)\ra_\n.
\]
After using Wick contraction and discarding the purely divergent terms\footnote{This is similar to the case of the free bosons in the previous section.}, we obtain
\[
\Z^{(2)}_\n=&\Z^{(0)}_\n\int_{\T_1}\int_{\text{T}_2^2}\la \L^{(1)}(z_1,\zb_1)\L^{(1)}(z_2,\zb_2)\ra_{\n}\nn\\
=&\frac{1}{\pi^4}\Z_\n^{(0)}\int_{\T_1}\int_{\T_2}\la T\Tbar^{(0)}(z_1,\zb_1) T\Tbar^{(0)}(z_2,\zb_2)\ra_\n\nn\\
=&\frac{1}{\pi^4}\Z_\n^{(0)}\int_{\T_1}\int_{\T_2}\Bigg\{\frac{1}{4}\left\lvert\en\right\rvert^4+\frac{1}{4}\left\lvert\pz\pn(\zer)\right\rvert^4+\frac{1}{4}\left\lvert\pn(\zer)\pz^2\pn(\zer)\right\rvert^2\nn\\
-&\frac{1}{4}\Big(\big(\pzb\pnb(\zber)\big)^2\pn(\zer)\pz^2\pn(\zer)+\big(\pz\pn(\zer)\big)^2\pnb(\zber)\pzb^2\pnb(\zber)\Big)\nn\\
+&\frac{1}{8}\Big(\en^2\pnb(\zber)\pdb^2\pnb(\zber)+\enbar^2\pn(\zer)\pd^2\pn(\zer)\Big)-\frac{1}{8}\Big(\en^2\big(\pzb\pnb(\zber)\big)^2+\enbar^2\big(\pz\pn(\zer)\big)^2\Big)\Bigg\}.\label{diracTTTT}
\]
The integrals of the nontrivial integrands shown above are listed below
\[
&\int_{\T_1}\int_{\T_2}\big(\pz\pn(\zer)\big)^2=\t_2\en\big(\pi-2\t_2\ete\big)+\t_2^2\big(\en^2-\frac{g_2}{6}\big),\label{6161}\\
&\int_{\T_1}\int_{\T_2}\pn(\zer)\pz^2\pn(\zer)=-\int_{\T_1}\int_{\T_2}\big(\pz\pn(\zer)\big)^2,\\
&\int_{\T_1}\int_{\T_2}\left\lvert\pz\pn(\zer)\right\rvert^4\nn\\
=&\t_2^2\left\lvert\en^2-\frac{g_2}{6}\right\rvert^2+\llv\en\rrv^2\big(4\t_2^2\llv\ete\rrv^2-2\pi\t_2(\ete+\eteb)\big)\nn\\
+&\Big(\t_2\en\big(\enbar^2-\frac{\bar{g}_2}{6}\big)(\pi-2\t_2\ete)+\t_2\enbar\big(\en^2-\frac{g_2}{6}\big)(\pi-2\t_2\eteb)\Big),\\
&\inter\llv\pn(\zer)\pz^2\pn(\zer)\rrv^2=\int_{\T_1}\int_{\T_2}\left\lvert\pz\pn(\zer)\right\rvert^4,\\
&\inter\big(\pzb\pnb(\zber)\big)^2\pn(\zer)\pd^2\pn(\zer)=-\int_{\T_1}\int_{\T_2}\llv(\pz\pn(\zer)\rrv^4.\label{6161fermion}
\]
For the detailed discussions of the above integrals please refer to Appendix \ref{b.2}.

With the help of the above nontrivial integrals and identity involving $g_2$, $\en$, and $\ete$
\[
g_2=6\big(\en^2-i\pi\pdt\en-2\ete\en\big),\label{g2two}
\]
one can find that \eqref{diracTTTT} equals
\[
&\frac{1}{\pi^4}\Z_\n^{(0)}\int_{\T_1}\int_{\T_2}\la T\Tbar^{(0)}(z_1,\zb_1) T\Tbar^{(0)}(z_2,\zb_2)\ra_\n\nn\\
=&\frac{\t_2^2}{2^4\pi^4}\llv\en\rrv^4+\frac{\t_2^2}{\pi^2}\t_2^2\llv\pdt\en\rrv^2+\frac{i\t_2^2}{4\pi^3}\big(\en^2\pdtb\enbar-\enbar^2\pdt\en\big)\nn\\
+&\frac{i\t_2}{\pi^2}\big(\enbar\pdt\en-\en\pdtb\enbar\big)-\frac{\t_2}{4\pi^3}\big(\en^2\enbar+\enbar^2\en\big)\nn\\
=&\frac{16}{\Z_\n^{(0)}}\big(\t_2^2\pdt^2\pdtb^2+i\t_2(\pdt^2\pdtb-\pdtb^2\pdt)\big)\Z_{\n}^{(0)}.
\]
Therefore the second-order corrections of the partition function with spin structure $\n$ are
\[
\Z_{\n}^{(2)}=&\frac{1}{\pi^4}\Z_\n^{(0)}\int_{\T_1}\int_{\T_2}\la T\Tbar^{(0)}(z_1,\zb_1) T\Tbar^{(0)}(z_2,\zb_2)\ra_\n\nn\\
=&16\big(\t_2^2\pdt^2\pdtb^2+i\t_2(\pdt^2\pdtb-\pdtb^2\pdt)\big)\Z_\n^{(0)},\label{z2dirac}
\]
which has the same structure with the bosonic case\eqref{boson2order}, and agrees with the result in \cite{Datta:2018thy}. Similar to the case of free bosons, for the deformed free Dirac fermions we can find the  counterterm corresponding to the minimum subtraction scheme as follows\footnote{The derivation is presented in Appendix \ref{sec-ct}.}
\[
\L_{\text{DF},\text{ct}}=\l^2\cdot\left\{\frac{8g^2}{\pi\e^2}\pd\psi^*\psi\pdb\bar{\psi}^*\bar{\psi}+\frac{1}{24\pi^3\e^6}\right\}.\label{FDFct}
\]

\section{Free Majorana fermions}\label{secMF}
As the last example, we investigate the deformation of  free massless Majorana {fermions}, whose un-deformed action is given by
\[
S=\frac{g}{2}\int_{\T}(\Psi^{T}\gamma^0\gamma^a\pd_a\Psi-\pd_a\Psi^T\gamma^0\gamma^a\Psi),
\]
where $\Psi$=$[\psi~~\bar{\psi}]^T$, the gamma matrices are defined in the previous section.

Similar to  the case of complex fermions, the $T\bar{T}$ flow of Lagrangian truncates at the first order, that is we have
\[
\L^\l=&\L^{(0)}+\l \L^{(1)},\quad T_{\m\n}^\l=T_{\m\n}^{(0)}+\l T_{\m\n}^{(1)},\label{Lmajor}
\]
where
\[
\L^{(0)}=2g\big(\pse\pzb\pse+\psr\pz\psr\big),\quad\L^{(1)}=&\frac{1}{\pi^2}\Big(\left(\Theta^{(0)}\right)^2-T^{(0)}\Tbar^{(0)}\Big)=g^2(2\pse\pzb\pse\psr\pz\psr-4\pse\pz\pse\psr\pzb\psr),\label{lmajor}
\]
and
\[
T^{(0)}=&-2\pi g\cdot\pse\pz\pse,\quad \Theta^{(0)}=-\pi g\cdot\big(\pse\pzb\pse+\psr\pz\psr\big),\quad \bar{T}^{(0)}=-2\pi g\cdot\psr\pzb\psr,\\
T^{(1)}=&2\pi g^2\cdot\pse\pz\pse\psr\pz\psr,\quad \Theta^{(1)}=0,\quad \bar{T}^{(1)}=2\pi g^2\cdot\psr\pzb\psr\pse\pzb\pse.\label{tmajor}
\]
Note that one could obtain (\ref{lmajor}--\ref{tmajor}) by simply removing the "$*$" in (\ref{l0dirac}--\ref{thetadirac}).

The un-deformed partition function with spin structure $\n$ is \cite{DiFrancesco:1997nk}
\[
\Z^{(0)}_\n=d_\n\bar{d}_\n,\quad d_\n(\t)=\left(\frac{\vartheta(\t)_\nu}{\eta(\t)}\right)^{1/2},
\]
The two-point functions for Majorana fermions with spin structure $\n$ are\cite{DiFrancesco:1997nk}
\[
\la\psi(z_1)\psi(z_2)\ra_\n=&(4\pi g)^{-1}\pn(\zer),\\
\la\bar{\psi}(\zb_1)\bar{\psi}(\zb_2)\ra=&(4\pi g)^{-1}\pnb(\zber),\\
\quad\text{others}=&0,\quad\n=2,3,4.
\]
Taking
derivatives on above propagators {gives}
\[
\la\pz\psi(z_1)\psi(z_2)\ra_\n=(4\pi g)^{-1}\pz\pn(\zer),\quad\la\pzb\psi(z_1)\psi(z_2)\ra_\n=(4 g)^{-1}\d^{(2)}(z_{12}).
\]
The regularized expectation value of the propagators and their derivatives when two points coincide are
\[
\la\psi(z_1)\psi(z_1)\ra_\n\equiv&\lim\limits_{z_2\rightarrow z_1}\big(\la\psi(z_1)\psi(z_2)\ra_\n-(4\pi g\zer)^{-1}\big)=0,\\
\la\pz\psi(z_1)\psi(z_1)\ra_\n\equiv&\lim\limits_{z_2\rightarrow z_1}\big(\la\pz\psi(z_1)\psi(z_2)\ra_\n+(4\pi g\zer^2)^{-1}\big)=-(8\pi g)^{-1}\en.\\
\la\pzb\psi(z_1)\psi(z_1)\ra_\n\equiv&\lim\limits_{z_2\rightarrow z_1}\big(\la\pzb\psi(z_1)\psi(z_2)\ra_\n-(4g)^{-1}\d(\zer)\big)=0,\\
\la\pzb\psi(z_1)\pz\psi(z_1)\ra_\n\equiv&\lim\limits_{z_2\rightarrow z_1}\big(\la\pzb\psi(z_1)\pz\psi(z_2)\ra_\n+(4g)^{-1}\pd\d(\zer)\big)=0,\\
\la\pzb\psi(z_1)\pzb\psi(z_1)\ra_\n\equiv&\lim\limits_{z_2\rightarrow z_1}\big(\la\pzb\psi(z_1)\pzb\psi(z_2)\ra_\n+(4g)^{-1}\pzb\d(\zer)\big)=0.
\]
{In analogy to the Dirac fermion case} we now go on to compute the corrections to the partition function.
\subsection{First-order}
 According to \eqref{lmajor}, the first-order correction of the partition function is
 \[
 \Z^{(1)}_\n=\frac{1}{\pi^2}\Z^{(0)}_\n\int_{\T}\left(T\Tbar^{(0)}(z,\zb)-(\Theta^{(0)})^2(z,\zb)\right)=&4g^2\t_2\Z_\n^{(0)}\la\pse\pz\pse\psr\pzb\psr\ra-g\t_2\Z_{\n}^{(0)}\la\pse\pzb\pse\psr\pz\psr\ra\nn\\
 =&\frac{\t_2}{(4\pi)^2}\Z_{\n}^{(0)}\llv\en\rrv^2\nn\\
 =&4\t_2\pdt\pdtb \Z^{(0)}_\n,\label{Major1order}
 \]
{which takes the same form that of in} free massless bosons and free massless Dirac fermions.
\subsection{Second-order}
For the second-order correction, in full analogy with  the case of Dirac fermions, there is no contribution that comes from $\la(\Theta^{(0)})^2(z_1,\zb_1)(\Theta^{(0)})^2(z_2,\zb_2)\ra$ and\\ $\la T\Tbar^{(0)}(z_1,\zb_1)(\Theta^{(0)})^2(z_2,\zb_2)\ra$, hence we go on to compute the only nonzero contribution $\la T\Tbar^{(0)}(z_1,\zb_1) T\Tbar^{(0)}(z_2,\zb_2)\ra$ and its integral.
\[
&\la T\Tbar^{(0)}(z_1,\zb_1)T\Tbar^{(0)}(z_2,\zb_2)\ra\nn\\
=&(2\pi g)^4\la\pse(z_1)\pz\pse(z_1)\bar{\pse}(z_1)\pzb\bar{\pse}(z_1)\pse(z_2)\pz\pse(z_2)\bar{\pse}(z_2)\pzb\bar{\pse}(z_2)\ra\nn\\
=&\frac{1}{16}\Bigg\{\frac{1}{16}\llv\en\rrv^4+\llv\pz\pn(\zer)\rrv^4+\llv\pn(\zer)\pz^2\pn(\zer)\rrv^2\nn\\
-&\Big(\big(\pzb\pnb(\zber)\big)^2\pn(\zer)\pz^2\pn(\zer)+\big(\pz\pn(\zer)\big)^2\pnb(\zber)\pzb^2\pnb(\zber)\Big)\nn\\
+&\frac{1}{4}\Big(\en^2\pnb(\zber)\pdb^2\pnb(\zber)+\enbar^2\pn(\zer)\pd^2\pn(\zer)\Big)-\frac{1}{4}\Big(\en^2\big(\pzb\pnb(\zber)\big)^2+\enbar^2\big(\pz\pn(\zer)\big)^2\Big)\Bigg\}.\label{8333}
\]
Utilizing the nontrivial integrals and the identity (\ref{6161})--(\ref{g2two}) mentioned before, the double integral of (\ref{8333}) equals
\[
&\inter\la T\Tbar^{(0)}(z_1,\zb_1)T\Tbar^{(0)}(z_2,\zb_2)\ra\nn\\
=&\frac{\t_2^2}{4^4}\llv\en\rrv^4+\frac{\pi^2\t_2^2}{4}\llv\pdt\en\rrv^2+\frac{i\pi\t_2^2}{32}\big(\en^2\pdtb\enbar-\enbar^2\pdt\en\big)\nn\\
-&\frac{\pi\t_2}{32}\big(\en^2\enbar+\enbar^2\en\big)-\frac{i\pi^2\t_2}{4}\big(\en\pdtb\enbar-\enbar\pdt\en\big)\nn\\
=&\frac{16\pi^4}{\Z_{\n}^{(0)}}\Big(\t_2^2\pdt^2\pdtb^2+i\t_2\big(\pdt^2\pdtb-\pdtb^2\pdt\big)\Big)\Z_{\n}^{(0)}\label{8444}.
\]

According to \eqref{8444}, we can obtain that the second-order correction of the partition function for deformed free Majorana fermions
\[
\Z^{(2)}_{\n}=&\frac{1}{\pi^4}\Z_\n^{(0)}\int_{\T_1}\int_{\T_2}\la T\Tbar^{(0)}(z_1,\zb_1) T\Tbar^{(0)}(z_2,\zb_2)\ra_\n\nn\\
=&16\big(\t_2^2\pdt^2\pdtb^2+i\t_2(\pdt^2\pdtb-\pdtb^2\pdt)\big)\Z_\n^{(0)},\label{Major2order}
\]
as was expected, the second-order corrections of Majorana fermions share the same structure as Dirac fermions \eqref{z2dirac} and free bosons \eqref{boson2order}, the conclusion of ref.\cite{Datta:2018thy} is confirmed again. Once again the  counterterm can be found as\footnote{The derivation is presented in Appendix \ref{sec-ct}.}
\[
\L_{\text{MF},\text{ct}}=\l^2\cdot\left\{\frac{8g^2}{\pi\e^2}\pd\psi\psi\pdb\bar{\psi}\bar{\psi}+\frac{1}{96\pi^3\e^6}\right\}.\label{FMFct}
\]

It's natural to ask whether the counterterms Eq.\eqref{FDFct}\eqref{FMFct} introduced in the first two orders are enough or not to cancel the divergences of the higher-order partition function in free fermionic theories.  From the perspective of the $T\bar{T}$ deformation as a kind of irrelevant deformation, one can expect new divergent terms to appear in the higher-order, and there is no a priori reason that new counterterms added to canceling these divergences should be vanishing, although the higher-order deformations of the Lagrangians \eqref{truncation}\eqref{lmajor} are truncated due to the Grassmannian structure of the fermion.  It is an interesting future problem to perform higher-order calculations to determine the exact higher-order counterterms.
\section{The first KdV charge}\label{secKDV}
In the previous sections, the corrections of various $T\bar{T}$ deformed partition functions evaluated by the conformal perturbation theory based on Lagrangian path integral are in good agreement with results obtained by the non-perturbative approach\cite{Datta:2018thy}. In this section, we proceed with the perturbation method to study the $T\bar{T}$-flow effects of the first quantum KdV charge\footnote{We are grateful to the anonymous referee's suggestion to study the $T\bar{T}$ deformation of KdV charge.}\cite{Bazhanov:1994ft},  for which there have been  studies based on non-perturbation methods\cite{LeFloch:2019wlf,Asrat:2020jsh}.

Let's first consider a generic CFT, for the sake of convenience we call it a seed later, on a cylinder with coordinate $\{z,\zb\}$ and circumference $L$\footnote{$z=x+iy,~\zb=x-iy,~x\sim x+L$.}. After the $T\bar{T}$ deformation, the deformed left-moving KdV charges $P^\l_s$ in the resulting QFT take the form
\[
P_s^\l=\frac{1}{2\pi}\int_0^L\big(\dd zT^\l_{s+1}+\dd\zb \Theta^\l_{s-1}\big),
\]
where the superscript $\l$ represents the deformation parameter. For $s=1$
\[
P^\l_1=\frac{1}{2\pi}\int_0^L\big(\dd zT^\l+\dd\zb \Theta^\l\big)\equiv-\frac{H^\l+P^\l}{2},\label{p1}
\]
where $H^\l=-\int_0^L\dd xT^\l_{yy}$ is deformed Hamiltonian and $P^\l=-i\int_{0}^{L}\dd xT^\l_{xy}$ is deformed momentum. The  expectation value of $P_1^\l$ in the deformed state $\lvert n\ra^\l$ thus reads
\[
^\l\la n\lvert P^\l_1\lvert n\ra^\l=-\frac{\mathcal{E}^\l_n+P^\l_n}{2},
\]
where $\mathcal{E}^\l_n$ and $P_n^\l$ represent the energy and momentum of the state $\lvert n\ra^\l$ respectively. From the $T\bar{T}$-flow equations of $\mathcal{E}^\l_n$, $P^\l_n$\cite{Smirnov:2016lqw,Zamolodchikov:2004ce,Cavaglia:2016oda,Jorjadze:2020ili}
\[
\mathcal{E}^\l_n=\frac{L}{2\l}\left(\sqrt{1+\frac{4\l E_n}{L}+\frac{4\l^2(P_n)^2}{L^2}}-1\right),~~P^\l_n=P_n,
\]
where $E_n$ and $P_n$ are energy and momentum of the undeformed eigenstate $\lvert n\ra$ in the seed, we could get the closed form for $^\l\la n\lvert P^\l_1\lvert n\ra^\l$depended only on $E_n$, $P_n$, $\l$, and $L$
\[
^\l\la n\lvert P^\l_1\lvert n\ra^\l=\frac{L}{4\l}\left(1-\sqrt{1+\frac{4\l E_n}{L}+\frac{4\l^2(P_n)^2}{L^2}}\right)-\frac{P_n}{2}.
\]
From now on, we're going to focus on the case where $n=0$ (i.e., the ground state) and the seed theory is free bosons or free Dirac fermions or free Majorana fermions.
 \subsection{Non-perturbative approach}
 For the seed theory being free bosons with periodic boundary condition$\big(\phi(z+L)=\phi(z)\big)$, or Dirac and Majorana fermions with anti-periodic boundary condition$\big(\psi(z+L)=-\psi(z)\big)$, the vacuum energy and momentum are
 \[
 E_0=-\frac{\pi c}{6L},~~~~P_0=0,
 \]
respectively, which leads to $^\l\la 0\lvert P^\l_1\lvert 0\ra^\l$ equals\footnote{We denote $^\l\la0\lvert\mo^\l\lvert0\ra^\l$ as $\la\mo^\l\ra_\o^\l$ for any flowing operator $\mo^\l$ on a cylinder.}
\[
\la P^\l_1\ra_\o^{\l}\equiv~^\l\la0\lvert P_1^\l\lvert0\ra^\l=&\frac{L}{4\l}\left(1-\sqrt{1-\frac{2\l \pi c}{3L^2}}\right)\nn\\
=&\frac{c\pi}{12 L}+\l\cdot\frac{c^2\pi^2}{72L^3}+O(\l^2)\nn\\
=&\frac{c\pi}{12 }+\l\cdot\frac{c^2\pi^2}{72}+O(\l^2),~~(\text{for}~L=1).\label{antires}
\]
 For the Dirac or Majorana fermions with periodic boundary condition$\big(\psi(z+L)=\psi(z)\big),$  the vacuum energy and momentum are
 \[
 E_0=\frac{\pi c}{3L},~~~~P_0=0,
 \]
which leads to $\la P^\l_1\ra_\o^{\l}$ equals
\[
\la P^\l_1\ra_\o^{\l}=&\frac{L}{4\l}\left(1-\sqrt{1+\frac{4\l \pi c}{3L^2}}\right)\nn\\
=&-\frac{c\pi}{6 L}+\l\cdot\frac{c^2\pi^2}{18L^3}+O(\l^2)\nn\\
=&-\frac{c\pi}{6}+\l\cdot\frac{c^2\pi^2}{18}+O(\l^2),~~(\text{for}~L=1).\label{perires}
\]
We next to reproduce the above results(\ref{antires}, \ref{perires}) by utilizing conformal perturbative approach.
\subsection{Perturbative approach}
According to \eqref{p1}, computing $\la P^\l_1\ra_\o^\l$ amounts to compute the deformed one-point functions $\la T^\l \ra_\o^\l$ and $\la \Theta^\l\ra_\o^\l$. Thanks to our previous setup, we may obtain the one-point functions on cylinder  by taking the zero temperature limit of the corresponding one-point functions on torus, namely,
\[
\lim\limits_{\beta\rightarrow\infty}\la \mo^\l\ra^\l_{\tor.}\equiv\lim\limits_{\beta\rightarrow\infty}\left\{\text{Tr}\left[e^{-\beta H^\l}\right]^{-1}\cdot\text{Tr}\left[e^{-\beta H^\l}\mo^\l\right]\right\}=\la\mo^\l\ra_\o^\l.
\]
In Lagrangian path integral formalism, $\la \mo^\l\ra^\l_{\tor.}$  equals to
\[
&\la\mo^\l\ra^\l_{\tor.}\nn\\
=&\frac{1}{\Z^\l}\int\mathcal{D}\phi\mo^\l\exp\left\{-\int_{\text{T}^2}\L^\l\right\}\nn\\
=&\la\mo^{(0)}\ra_{\tor.}+\lambda\cdot\left\{\la\mo^{(1)}\ra_{\tor.}+\la\mo^{(0)}\ra_{\tor.}\int_{\text{T}^2}\la\L^{(1)}\ra_{\tor.}-\int_{\text{T}^2_1}\la\mo\L^{(1)}(z_1,\zb_1)\ra_{\tor.}\right\}+O(\l^2),\label{1pttor}
\]
where $\mo^\l=\sum_{n=0}^{\infty}\frac{\l^n}{n!}\mo^{(n)}$, $\la \mo\ra_{\tor.}\equiv\text{Tr}\left[e^{-\beta H}\right]^{-1}\cdot\text{Tr}\left[e^{-\beta H}\mo\right]$. We then make use of (\ref{1pttor}) to calculate the $\la T^\l\ra^\l_{\tor.}$ and $\la\Theta^\l\ra^\l_{\tor.}$ of free bosons and free fermions respectively.

With the help of free propagators given in previous sections, after doing Wick contraction and simple integral on a torus, the final results are listed as follows.
For free bosons, we get\footnote{For the definition of $\eta_1$ and $e_{\n-1}$, please refer to Appendix \ref{a}.}
\[
\la T^\l\ra^\l_{\tor.\FB}=&\left(\eta_1-\frac{\pi}{2\t_2}\right)+\lambda\cdot\left(\frac{2\lvert\eta_1\lvert^2}{\pi}-\frac{1}{2\t_2}\Big(\eta_1+\bar{\eta}_1\Big)+\Big(\frac{2}{\pi}\t_2\etb_1-1\Big)i\pdt\eta_1\right)+O(\l^2),\label{T1boson}\\
\la\Theta^\l\ra^\l_{\tor.\FB}=&\lambda\cdot\left(-\frac{\lvert\eta_1\lvert^2}{\pi}+\frac{2}{\t_2}\Big(\eta_1+\etb_1\Big)-\frac{3\pi}{4\t_2^2}\right)+O(\l^2).
\]
For free Dirac fermions,
\[
\la T^\l\ra^\l_{\n;\tor.\DF}=&-\frac{e_{\n-1}}{2}+\lambda\cdot\left(\frac{\lvert e_{\n-1}\lvert^2}{2\pi}+\frac{i\t_2\bar{e}_{\n-1}\pdt e_{\n-1}}{2\pi}\right)+O(\l^2),\\
\la\Theta^\l\ra^\l_{\n;\tor.\DF}=&\lambda\cdot\frac{-\lvert e_{\n-1}\lvert^2}{4\pi}+O(\l^2),~~~\n=2,3,4.
\]
And for free Majorana fermions,
\[
\la T^\l\ra^\l_{\n;\tor.\MF}=&-\frac{e_{\n-1}}{4}+\lambda\cdot\left(\frac{\lvert e_{\n-1}\lvert^2}{8\pi}+\frac{i\t_2\bar{e}_{\n-1}\pdt e_{\n-1}}{8\pi}\right)+O(\l^2),\\
\la\Theta^\l\ra^\l_{\n;\tor.\MF}=&\lambda\cdot\frac{-\lvert e_{\n-1}\lvert^2}{16\pi}+O(\l^2),~~~\n=2,3,4.\label{theta1major}
\]
Take the zero temperature limits of (\ref{T1boson}--\ref{theta1major}) respectively, one obtains\footnote{Here we set the modular parameter $\t=i\t_2=i\beta$.}
\[
\la T^\l\ra^\l_{\o,\FB}&=\lim\limits_{\t_2\rightarrow\infty}\la T^\l\ra^\l_{\tor.\FB}=\frac{\pi^2}{6}+\lambda\cdot\frac{\pi^3}{18}+O(\l^2),\label{fbt}\\
\la \Theta^\l\ra^\l_{\o,\FB}&=\lim\limits_{\t_2\rightarrow\infty}\la \Theta^\l\ra^\l_{\tor.\FB}=\lambda\cdot\frac{-\pi^3}{36}+O(\l^2),\label{fbtheta}
\]
\[
\la T^\l\ra^\l_{\o,\DF}&=\lim\limits_{\t_2\rightarrow\infty}\la T^\l\ra^\l_{\n;\tor.\DF}=
\begin{cases}
\frac{-\pi^2}{3}+\lambda\cdot\frac{2\pi^3}{9}+O(\l^2), & \n=2,\\
\frac{\pi^2}{6}+\l\cdot\frac{\pi^3}{18}+O(\l^2), &\n=3,4,
\end{cases}\\
\la \Theta^\l\ra^\l_{\o,\DF}&=\lim\limits_{\t_2\rightarrow\infty}\la \Theta^\l\ra^\l_{\n;\tor.\DF}=
\begin{cases}
\lambda\cdot\frac{-\pi^3}{9}+O(\l^2), & \n=2,\\
\l\cdot\frac{-\pi^3}{36}+O(\l^2), &\n=3,4,
\end{cases}
\]
\[
\la T^\l\ra^\l_{\o,\MF}&=\lim\limits_{\t_2\rightarrow\infty}\la T^\l\ra^\l_{\n;\tor.\MF}=
\begin{cases}
\frac{-\pi^2}{6}+\lambda\cdot\frac{\pi^3}{18}+O(\l^2), & \n=2,\\
\frac{\pi^2}{12}+\l\cdot\frac{\pi^3}{72}+O(\l^2), &\n=3,4,
\end{cases}\\
\la \Theta^\l\ra^\l_{\o,\MF}&=\lim\limits_{\t_2\rightarrow\infty}\la \Theta^\l\ra^\l_{\n;\tor.\MF}=
\begin{cases}
\lambda\cdot\frac{-\pi^3}{36}+O(\l^2), & \n=2,\\
\l\cdot\frac{-\pi^3}{144}+O(\l^2), &\n=3,4.
\end{cases}
\]
Note that for the fermion cases, $\n=2,3,4$  correspond to the periodic(space)-antiperiodic(time), antiperiodic-periodic and antiperiodic-antiperiodic sectors respectively. It means that for fermions on a cylinder of circumference unity with periodic B. C.
\[
\la T\ra^\l_{\o,\DF}=&\frac{-\pi^2}{3}+\l\cdot\frac{2\pi^3}{9}+O(\l^2),~~\la \Theta\ra^\l_{\o,\DF}=-\l\cdot\frac{\pi^3}{9}+O(\l^2),\label{DFT}\\
\la T\ra^\l_{\o,\MF}=&\frac{-\pi^2}{6}+\l\cdot\frac{\pi^3}{18}+O(\l^2),~~\la \Theta\ra^\l_{\o,\MF}=-\l\cdot\frac{\pi^3}{36}+O(\l^2),
\]
and
\[
\la T\ra^\l_{\o,\DF}=&\frac{\pi^2}{6}+\l\cdot\frac{\pi^3}{18}+O(\l^2),~~\la \Theta\ra^\l_{\o,\DF}=-\l\cdot\frac{\pi^3}{36}+O(\l^2),\\
\la T\ra^\l_{\o,\MF}=&\frac{\pi^2}{12}+\l\cdot\frac{\pi^3}{72}+O(\l^2),~~\la \Theta\ra^\l_{\o,\MF}=-\l\cdot\frac{\pi^3}{144}+O(\l^2),\label{MFtheta}
\]
for antiperiodic B.C.

With the help of \eqref{p1}, (\ref{fbt}--\ref{fbtheta}), and (\ref{DFT}--\ref{MFtheta}), we obtain the $T\bar{T}$-flow of the KdV charge $P_1$ up to the first-order for three free theories
\[
\la P^\l_1\ra^\l_{\o,\FB}=&\frac{\pi}{12}+\l\cdot\frac{\pi^2}{72}+O(\l^2),~~(\text{periodic B.C.})\label{p1fb}\\
\la P^\l_1\ra^\l_{\o,\DF}=&\frac{\pi}{12}+\l\cdot\frac{\pi^2}{72}+O(\l^2),~~(\text{antiperiodic B.C.})\\
\la P^\l_1\ra^\l_{\o,\MF}=&\frac{\pi}{24}+\l\cdot\frac{\pi^2}{288}+O(\l^2),~~(\text{antiperiodic B.C.})\label{p1mfp}
\]
\[
\la P^\l_1\ra^\l_{\o,\DF}=&-\frac{\pi}{6}+\l\cdot\frac{\pi^2}{18}+O(\l^2),~~(\text{periodic B.C.})\label{DFanti}\\
\la P^\l_1\ra^\l_{\o,\MF}=&-\frac{\pi}{12}+\l\cdot\frac{\pi^2}{72}+O(\l^2).~~(\text{periodic B.C.})\label{MFanti}
\]
It's easy to check that the perturbative results Eq.(\ref{p1fb}--\ref{p1mfp}) (Eq.(\ref{DFanti}--\ref{MFanti})) match the results come from non-perturbative method Eq.(\ref{antires}) (Eq.(\ref{perires}))\footnote{The center charges
 $c_{FB}=c_{DF}=2c_{MF}=1$.}. The discussions on whether these two approaches match each other at the second-order,  which is technically involved, will be served as our future work.
\section{Conclusion and Discussion}\label{smy}
In this work,  we perturbatively calculate the  flow effects of $T\bar{T}$ deformation on the torus partition functions and the VEV of the first KdV charge $P_1$ under the Lagrangian path integral formalism. In previous cases \cite{He:2019vzf,He:2019ahx,He:2020udl}, the authors have studied the correlation {functions} perturbatively up to the first-order deformation. Generally speaking, to evaluate the  correlation functions and higher ordered partition functions perturbatively,  the flow of stress tensor must be taken into consideration. As a preliminary study, we focus on the discussions of  free theories, including free bosons,  Dirac fermions and  Majorana fermions, where the  flow of stress tensor  can be constructed  explicitly \footnote{For the discussions of $T$ and $\bar{T}$ flow in generic CFTs, please refer to \cite{Cardy:2019qao,Caputa:2020lpa}.}. In terms of Wick  contraction, we first compute the first- (\ref{boson1ordr},\ref{Dira1order},\ref{Major1order}) and the second-order (\ref{boson2order},\ref{z2dirac},\ref{Major2order}) deformations to the partition functions, then we calculate a certain kind of 1-pt function (\ref{p1fb}--\ref{MFanti}), i.e. VEV of the first KdV charge. It turns out that the first two orders of the deformed partition functions and the first-order VEV of the first quantum KdV charge  are in good agreement with the results obtained in Hamiltonian formalism \cite{Datta:2018thy}\cite{Asrat:2020jsh} respectively, provided we make minimum subtraction in dealing with the divergence.

Although the results obtained from the Hamiltonian formalism are reproduced in the Lagrangian path integral formalism, in general, due to the emergence of higher derivative terms in the deformed Lagrangian (\ref{bsh1}, \ref{bsh2}), the equivalence between the Lagrangian path integral formalism and the Hamiltonian path integral formalism remains as a mystery. For instance, though Legendre transformation, it can be found that the  Minkowski Hamiltonian of the deformed free bosons takes the form \cite{Kraus:2018xrn, Chakrabarti:2020pxr}
\[
\mathcal{H^\l}=&\frac{1}{2\l}\Big(-1+\sqrt{1+2\l\big(\pi^2+\phi'^2\big)+4\l^2(\pi\phi')^2}\Big)\nn\\
=&\frac{1}{2}\big(\pi^2+\phi'^2\big)+\frac{\l}{4}\big(-\pi^4+2\pi^2\phi'^2-\phi'^4\big)+\frac{\l^2}{4}\big(\pi^6-\pi^4\phi'^2-\pi^2\phi'^4+\phi'^6\big)+\mo(\l^3),\label{9595}
\]
 where $\phi'$ is the spatial derivative of $\phi$ and $\pi$ the canonical momentum conjugate to $\phi$. The higher power terms of $\pi$ presented in \eqref{9595} prevent us from getting the Lagrangian path integral directly from the corresponding Hamiltonian path integral, since how to deal with the generic integrals go beyond Gaussian integrals, for now, is still a major problem for mathematicians and physicists. It also leads to an open question of whether the Hamiltonian formalism is more fundamental than the Lagrangian formalism\cite{feynman2012quantum, Peskin:1995ev, Kauffmann:2009wg, Erich:2013:Online}. Fortunately, our results show, for the $T\bar{T}$ deformed theory, the use of disk regularization \cite{Dijkgraaf:1996iy}  together with minimum subtraction in Lagrangian formalism seems to be sufficient to match the Hamiltonian formalism. For the instances considered in this paper, the second-order Lagrangian counterterms corresponding to the minimum subtraction are presented in (\ref{FBct}, \ref{FDFct}, \ref{FMFct}) respectively. To match the partition functions and correlation functions between the Lagrangian formalism and the Hamiltonian formalism up to the higher-order deformations will be interesting future work.

Further, it will be interesting to study the second-order deformation to the partition function in the interacting theories, e.g. massive fermions and bosons, Liouville field theory\cite{Leoni:2020rof}, and so on. The generic correlation functions with the $T\Tbar$-flow effects in SUSY extended CFTs will be also an interesting future direction with the following \cite{He:2019ahx}.

\subsection*{Acknowledgements}
We would like to thank Giulio Bonelli, Bin Chen, Pak Hang Chris Lau, Yi Li, Hao Ouyang, Hongfei Shu, and Stefan Theisen for useful discussion. We are also grateful to anonymous referees for their constructive suggestions, which improve this manuscript significantly. S.H. would like to appreciate the financial support from Jilin University, Max Planck Partner group as well as Natural Science Foundation of China Grants (No.12075101, No. 1204756).
Y.S. would like to appreciate the support from China Postdoctoral Science Foundation (No. 2019M653137).
\appendix
\section{Details of Weierstrass functions}\label{a}
In this Appendix, we give the definitions and properties of  Weierstrass functions that appear in the calculations.

We first note that, in our convention, torus ($\text{T}^2$) is defined by the identification on complex plane $z\sim z+2w$ and $z\sim z+2w'$ with $2w=1$, $2w'=\t=\t_1+i\t_2$.

The first Weierstrass function $P(z)$, called \textit{Weierstrass P-function}, is defined as
\[
P(z)=\frac{1}{z^2}+\sum_{\{m,n\}\neq\{0,0\}}\Big(\frac{1}{(z-\tilde{w})^2}-\frac{1}{\tilde{w}^2}\Big),\quad\tilde{w}=2mw+2nw'.
\]
The Laurent series expansion of $P(z)$ in the neighborhood of $z=0$ is
\[
P(z)\sim \frac{1}{z^2}+\frac{g_2}{20}z^2+\frac{g_3}{28}z^4+\mo(z^6),
\]
hence we have
\[
\pd P(z)\sim&-\frac{2}{z^3}+\frac{g_2}{10}z+\frac{g_3}{7}z^3+\mo(z^5),\quad\pd^2P(z)\sim\frac{6}{z^4}+\frac{g_2}{10}+\frac{3g_3}{7}z^2+\mo(z^4),
\]
where $g_2$ and $g_3$ are called \textit{Weierstrass Invariants}
\[
g_2:=\sum_{\{m,n\}\neq\{0,0\}}\frac{60}{\tilde{w}^4},\quad g_3:=\sum_{\{m,n\}\neq\{0,0\}}\frac{140}{\tilde{w}^6}.
\]
The second Weierstrass function $\zeta(z)$, called \textit{Weierstrass zeta-function}, is a primitive function of $-P(z)$
\[
\zeta(z)=&\frac{1}{z}+\sum_{\{m,n\}\neq\{0,0\}}\Big(\frac{1}{z-\tilde{w}}+\frac{1}{\tilde{w}}+\frac{z}{\tilde{w}^2}\Big),\quad\pd\zeta(z)=-P(z).
\]
We then define
\[
\ete:=\zeta(w),\quad\eta_2:=\zeta(w'),
\]
and
\[
e_1:=P(w),~~e_2:=P(-w-w'),~~e_3:=P(w'),
\]
which are functions of the modular parameter $\t$.  Note that there is an identity about $\ete(\t)$ and Dedekind eta function $\eta(\t)$,
%\footnote{We will take this opportunity to point our several mistakes in previous work \cite{He:2020udl} by two of us. In appendix A of \cite{He:2020udl}, we identify $\eta(\tau)$ with $\eta_1$, %which is not the case. Thus (42) in \cite{He:2020udl} should be replaced by (\ref{eta1}) here. The mistake doesn't effect our final conclusion. The second mistake is that (D24) in \cite{He:2020udl} %is wrong, since $\ln \s(z)$ is multi-valued when $z$ moves around $z=0$, thus the Stoke's theorem does not applicable, one way to solve this problem is by making use of (3.11) in  %\cite{Datta:2014zpa} (see also \cite{Dijkgraaf:1996iy}) where the same regularization method is used. It turns out (D25) in \cite{He:2020udl} will be
%\[
%B(a)-B(b)=2\tau_2 \eta_1(b-a)-\pi(b-a-\bar{b}+\bar{a})\nn
%\]
%and (D26) changes accordingly. }
\[\label{eta1}
\frac{\pdt\eta}{\eta}=\frac{i}{2\pi}\ete.
\]
which has been used in the  bosonic calculations \eqref{bonson1}.
\section{Details of some integrations}
\subsection{Prescription for regularization}\label{b.0}
Since the integrands over {a} torus we are interested in may contain singularities, in this Appendix we will discuss  how to deal with these singularities  based on the prescription given in \cite{Dijkgraaf:1996iy}.

Let us consider an integrand $f(z,\zb)$ defined on a torus, which contains $N$ number of singularities $(r_1,r_2...r_N)$.
Following the prescription in \cite{Dijkgraaf:1996iy}, {when performing the integrals, we integrate over not the whole torus T$^2$}, but over the regularized parallelogram---the parallelogram with small {disks} around the singularities removed (see Fig.\ref{fig1} for example). In the following, we  denote the regularized torus by $\text{T}'^2$.
\begin{figure}[htb]
\centering
\includegraphics[width=0.6\textwidth]{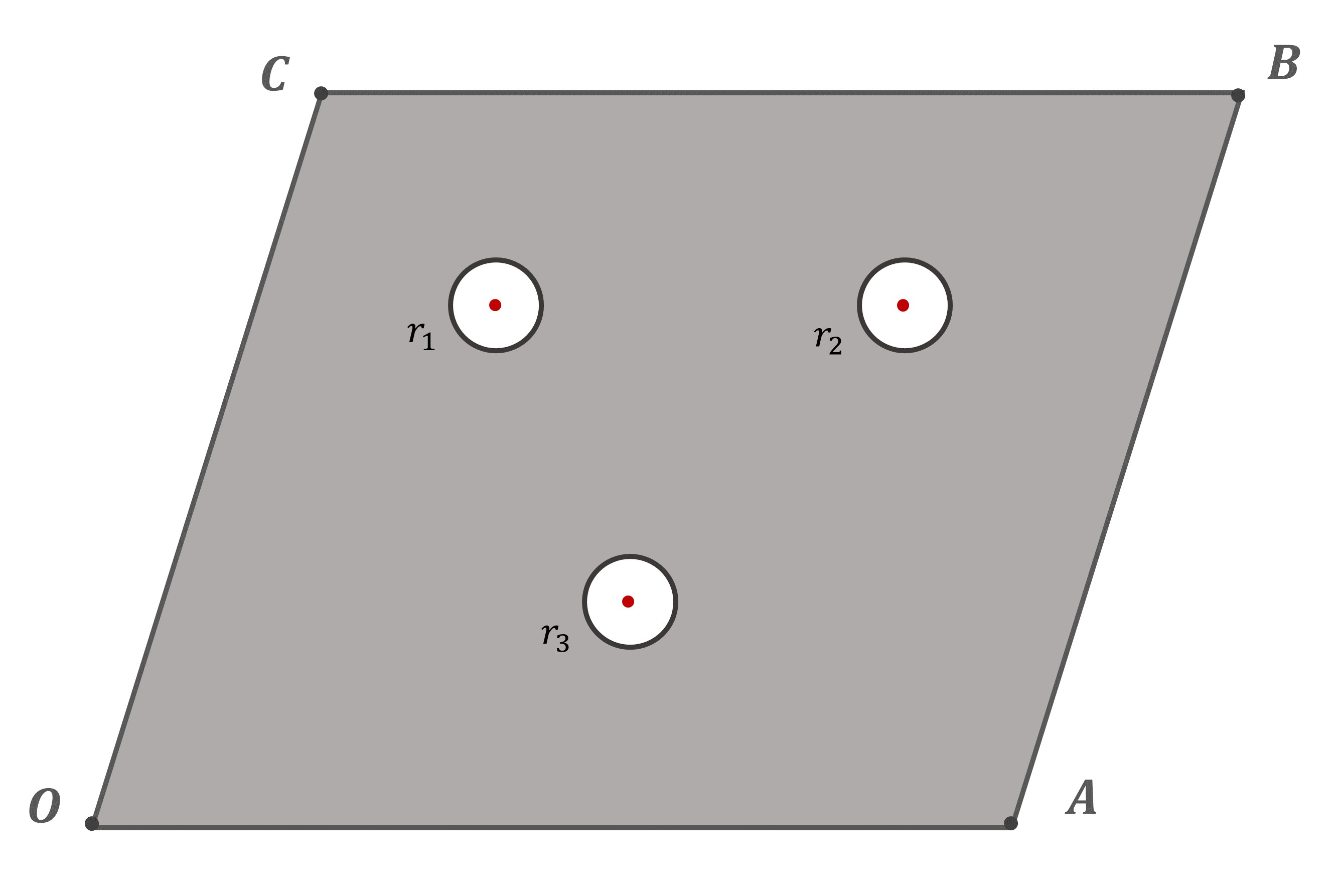}\\
\caption{The regularized cell for $f(z,\zb)$ contains three singularities (red points). The grey part bounded by the solid lines is the regularized integral region.}\label{fig1}
\end{figure}
Suppose we find that
\[
f(z,\zb)=\pd_\m F^\m(z,\zb),
\]
then with the Stoke's theorem in 2D space \footnote{Since $z=x+iy$, $\int_{\Sigma}\dd^2x\equiv\int_{\Sigma}\dd x\wedge\dd y=\frac{i}{2}\int_{\Sigma}\dd z\wedge\dd\zb\equiv\frac{i}{2}\int_{\Sigma}\dd^2z$.}
\[
\int_{\Sigma}f(z,\zb)\dd^2x=\frac{i}{2}\oint_{\pd\Sigma}\big(F^z\dd\zb-F^{\zb}\dd z\big),
\]
which can be applied to the regularized torus leading to
\[
\int_{\text{T}'^2}f(z,\zb)\dd^2x=\frac{i}{2}\Big[\oint_{\pd\text{T}^2}-\oint_{\pd\text{D}(\text{poles})}\Big]\big(F^z\dd\bar{z}-F^{\bar{z}}\dd z\big),
\]
where the contour integrals are anticlockwise. In this paper, we focus further on the case that $F^\m(z,\zb)$ can be  written as $F^\m(z,\bar{z})=F^\m_1(z)F^\m_2(\bar{z})$, where $F^\m_1$ is  holomorphic function and $F^\m_{2}$ is anti--holomorphic. For the $j$-th pole $(r_j,\bar{r}_j)$ of $f(z,\zb)$ in $\text{T}^2$, $F^\m(z,\zb)$ could be expanded around it as follows
\[
F^\m(z,\bar{z})=&\sum_{m}\sum_{n}C^{1,\m}_{j,m}C^{2,\m}_{j,n}(z-r_j)^m(\zb-\bar{r}_j)^n,
\]
then
\[
\oint_{|z-r_j|=\e}\big(F^z\dd\zb-F^{\zb}\dd z\big)=&\int_{0}^{2\pi}\sum_{m}\sum_{n}C^{1,z}_{j,m}C^{2,z}_{j,n}(\e e^{i\theta})^m(\e e^{-i\theta})^n(-i\e )e^{-i\theta}\dd\theta\nn\\
&-\int_{0}^{2\pi}\sum_m\sum_nC^{1,\zb}_{j,m}C^{2,\zb}_{j,n}(\e e^{i\theta})^m(\e e^{-i\theta})^n(i\e)e^{i\theta}\dd\theta\nn\\
=&-2\pi i\sum_{n}\e^{2(n+1)}\Big(C^{1,\zb}_{j,n}C^{2,\zb}_{j,n+1}+C^{1,z}_{j,n+1}C^{2,z}_{j,n}\Big).
\]
Therefore, on the grounds of the prescription in \cite{Dijkgraaf:1996iy}, we have
\[
\int_{\text{T}^2}f(z,\zb)\dd^2x:=&\int_{\text{T}'^2}f(z,\zb)\dd^2x\nn\\
=&\limr G(\e)+\frac{i}{2}\oint_{\pd\text{T}^2}\big(F^{z}\dd\zb-F^{\zb}\dd z\big),
\]
where
\[
G(\e):=-\pi\sum_{j,n}\e^{2(n+1)}\Big(C^{1,\zb}_{j,n}C^{2,\zb}_{j,n+1}+C^{1,z}_{j,n+1}C^{2,z}_{j,n}\Big).
\]
It is worth noting that for the case of $F^z$ holomorphic, meanwhile, $F^{\zb}$ anti-holomorphic, it must have $\lim\limits_{\e\rightarrow0}G(\e)=0$.
\subsection{Integrals for bosonic fields}\label{b.1}
In this Appendix we record the details of integrals appearing in the calculations of free {bosons} part (\ref{pms}--\ref{ps}).

Since all the integrands are double periodic, we can shift the variable of the integration to make life easier without changing the value of the integrals, i.e., $\inter f(z_{12},\zb_{12})=\t_2\int_{\T}f(z,\zb)$ for double periodic function $f$.

We start with the integration of the $P$-function in a cell. Since $P(z)=-\frac{\pd\zeta(z)}{\pd z}$, with the integral strategy shown in Appendix \ref{b.0}, we have \footnote{In this case $\limr G(\e)=0$.}
\[
\intt P(z)&=-\intt\frac{\pd\zeta(z)}{\pd z}=\frac{-i}{2}\oint_{\pd\text{T}'^2}\zeta(z)\dd\zb\nn\\
&=-\frac{i}{2}\Big(\oa+\ab+\bc+\co\Big)\zeta(z)\dd\zb \nn\\
&=-\frac{i}{2}\oa\big(\zeta(z)-\zeta(z+2w')\big)+\frac{i}{2}\int_{z_0}^{z_0+2w'}\dd\bar{z}\big(\zeta(z)-\zeta(z+2w)\big)\nn\\
&=2i\bar{w}\zeta(w')-2i\bar{w}'\zeta(w)=\pi-2\t_2\eta_1,\label{pz}
\]
where we have used the identity
\[
w'\zeta(w)-w\zeta(w')=\frac{i\pi}{2}
\]
to eliminate $\zeta(w')$.

Next Let us consider $\intt P(z)^2$. Since $P(z)^2$ is still a double periodic meromorphic function, we can expand $P(z)^2$ in terms of $\zeta(z)$ and its derivatives \cite{book:1157405},
\[
P(z)^2=\frac{g_2}{12}-\frac{1}{6}\zeta^{(3)}(z),\label{psquare}
\]
where the constant $\frac{g_2}{12}$ is fixed  by comparing the constant terms of Laurent expansion of two functions, $P(z)^2$ and $\zeta^{(3)}(z)$, at zero. Then
\[
\intt P(z)^2&=\frac{g_2}{12}\t_2-\frac{1}{6}\intt\zeta^{(3)}(z)\nn\\
&=\frac{g_2}{12}\t_2-\frac{1}{6}\intt\frac{\pd\zeta^{(2)}(z)}{\pd z}=\frac{g_2\t_2}{12}.\label{pzsquare}
\]
We next turn to the integrand $\llv P(z)\rrv^2$. Since $\llv P(z)\rrv^2$ is no longer analytic, we can not expand it in terms of $\zeta(z)$ as what we did for $P(z)^2$. Instead, we will adopt the following approach\footnote{Here we have omitted the term $\intt\zeta(z)\pd\bar{P}(\zb)$, since for any holomorphic function $f(z)$ with  poles $r_i$ of order $n_i$  and anti-holomorphic function $\bar{g}(\zb)$ with  poles $\bar{p}_k$ of order $m_k$
\[
\intt f(z)\pd\bar{g}(\zb)\sim&\intt f(z)\sum_{k}(\bar{\pd}^{m_k-1}\d^{(2)}(z-p_k))\nn\\
\sim&\sum_{k}\int_{T^2}\dd^2x(\pdb^{m_k-1}f(z))\delta^{(2)}(z-p_k)\nn\\
\sim&\sum_{i,k}\int_{T^2}\dd^2x(\pd^{n_i-1}\pdb^{m_k-2}\d^{(2)}(z-r_i))\delta^{(2)}(z-p_k)\nn\\
\sim&\sum_{i,k}\pd^{n_i-1}\pdb^{m_k-2}\d^{(2)}(p_k-r_i).
\]
For $f(z)=\zeta(z)$ and $\bar{P}(\zb)=\bar{g}(\zb)$, the result is purely divergent thus we have dropped it. We have discarded the similar terms in later integrals.}
\[
\intt P(z)\pb(\zb)=&-\int_{\text{T}'^2}\dd^2x\pd(\zeta(z)\pb(\zb))\nn\\
=&-\limgr-\frac{i}{2}\oint_{\pd\tf}\dd\zb\zeta(z)\pb(\zb)\nn\\
=&\limr\pi \e^{-2}-\frac{i}{2}\Big(\oa+\ab+\bc+\co\Big)\zeta(z)\bar{P}(\zb)\dd\zb\nn\\
=&2i(\eta_1\etb_2-\eta_2\etb_1)+\limr\pi \e^{-2}=4\tau_2|\ete|^2-2\pi(\ete +\eteb)+\limr\pi \e^{-2}.
\]
Note that the integration is divergent, which is consistent with the intuitive expectation to the integral process, since $\llv P(z)\rrv^2\sim\frac{1}{\llv z\rrv^4}$ when $z$ close to zero.
We regularize it by simply subtracting the divergent part, that is, we set
\footnote{In plane case, there is a similar divergence, which is moved out by dimensional regularization \cite{He:2019ahx}.}
\[
\intt \llv P(z)\rrv^2=4\tau_2|\ete|^2-2\pi(\ete +\eteb).\label{pzmode}
\]
Next consider the integrand
$P(z)^2\pb(\zb)=\frac{g_2}{12}\pb(\zb)-\frac{1}{6}\zeta^{(3)}(z)\pb(\zb)$, where we can use \eqref{psquare} to rewrite it as follows
\[
\intt P(z)^2\pb(\zb)=&\frac{g_2}{12}\intt\pb(\zb)-\frac{1}{6}\intt\zeta^{(3)}(z)\pb(\zb)\nn\\
=&\frac{g_2}{12}(\pi-2\eteb\t_2)-\frac{1}{6}\intt\pd\big(\zeta^{(2)}(z)\bar{P}(\zb)\big)\nn\\
=&\frac{g_2}{12}(\pi-2\eteb\t_2)-\frac{i}{12}\oint_{\pd\T}\zeta^{(2)}(z)\bar{P}(\zb)\dd\zb-\frac{1}{6}\limr G(\e)\nn\\
=&\frac{g_2}{12}(\pi-2\eteb\t_2).\label{psquarepbar}
\]
Finally, let us consider integration of $\llv P(z)\rrv^4=\big(\frac{g_2}{12}-\frac{1}{6}\zeta^{(3)}(z)\big)\big(\frac{\bar{g}_2}{12}-\frac{1}{6}\zetab^{(3)}(\zb)\big)$,
\[
&\intt P(z)^2\pb(\zb)^2\nn\\
=&\intt\big(\frac{g_2}{12}-\frac{1}{6}\zeta^{(3)}(z)\big)\big(\frac{\bar{g}_2}{12}-\frac{1}{6}\zetab^{(3)}(\zb)\big)\nn\\
=&\llv\frac{g_2}{12}\rrv^2\t_2-\frac{g_2}{72}\intt\bar{\zeta}^{(3)}(\zb)-\frac{\bar{g}_2}{72}\intt\zeta^{(3)}(z)+\frac{1}{36}\intt \zeta^{(3)}(z)\zetab^{(3)}(\zb)\nn\\
=&\frac{|g_2|^2\t_2}{12^2}+\frac{1}{36}\intt\pd\big(\zeta^{(2)}(z)\bar{\zeta}^{(3)}(\zb)\big)\nn\\
=&\frac{|g_2|^2\t_2}{12^2}+\frac{i}{72}\oint_{\pd\T}\zeta^{(2)}(z)\bar{\zeta}^{(3)}(\zb)\dd\zb+\frac{1}{36}\limr G(\e)\nn\\
=&\frac{|g_2|^2\t_2}{12^2}+\limr\frac{\pi}{3\e^6}.
\]
Similar to the case \eqref{pzmode}, we regularize the integral by simply discarding the divergent part, which gives
\[
\intt\llv P(z)\rrv^4=\frac{\llv g_2\rrv^2\t_2}{12^2}.\label{pzmodesquare}
\]
According to the results of \eqref{pz}, \eqref{pzsquare}, \eqref{pzmode}, \eqref{psquarepbar} and \eqref{pzmodesquare}, we have
\[
\int_{\T_1}\int_{\T_2}\big(B-P(z_{12})\big)=&\t_2^2\big(\frac{\pi}{\t_2}-2\ete\big)-\t_2\big(\pi-2\t_2\ete\big)=0,\\
\int_{\T_1}\int_{\T_2}\big(B-P(z_{12})\big)^2=&\t_2\intt\big(B^2+P(z)^2-2BP(z)\big)=\frac{g_2\t_2^2}{12}-\t_2^2B^2,\\
\int_{\T_1}\int_{\T_2}\llv B-P(z_{12})\rrv^2=&\t_2\intt\big(\llv B\rrv^2-B\bar{P}(\zb)-\bar{B}P(z)+\llv P(z)\rrv^2\big)=-\pi^2,
\]
and
\[
&\int_{\T_1}\int_{\T_2}\llv B-P(z_{12})\rrv^4\nn\\
=&\nn\t_2\intt\Big(\llv B\rrv^4+\llv P(z)\rrv^4+4\llv B\rrv^2\llv P(z)\rrv^2+\big(B^2\bar{P}(\zb)^2+\bar{B}^2P(z)^2\big)\nn\\
&-2\llv B\rrv^2\big(B\bar{P}(\zb)+\bar{B}P(z)\big)\Big)-2\big(BP(z)\bar{P}(\zb)^2+\bar{B}\bar{P}(z)P(z)^2\big)\Big)\nn\\
=&\t_2^2|B|^4+\frac{|g_2|^2\t_2^2}{12^2}-4\t_2^2A^2|B|^2-B^2\frac{\bar{g}_2\t_2^2}{12}-\bar{B}^2\frac{g_2\t_2^2}{12}.
\]
\subsection{Integrals for fermionic fields}\label{b.2}
In this Appendix we present the details of integrals appearing in the calculations of free {fermions} part (\ref{6161}--\ref{6161fermion}).

We first note that both $\big(\pd\pn(z)\big)^2$ and $\pn(z)\pd^2\pn(z)$ are elliptic functions with the modular parameter $\t$,
\[
\big(\pd\pn(z)\big)^2=\frac{\big(\pd P(z)\big)^2}{4\big(P(z)-\en\big)},\quad\pn(z)\pd^2\pn(z)=\frac{1}{2}\pd^2P(z)-\frac{\big(\pd P(z)\big)^2}{4\big(P(z)-\en\big)},
\]
where $e_1:=P(w)$, $e_2:=P(w+w')$, $e_3:=P(w')$.  Hence we can expand $\big(\pd\pn(z)\big)^2$ and $\pn(z)\pd^2\pn(z)$ in terms of $\zeta(z)$ and its derivatives, the results are
\[
\big(\pd\pn(z)\big)^2=&\frac{1}{6}\pd^2P(z)+\en P(z)+\en^2-\frac{g_2}{6},\\
\pn(z)\pd^2\pn(z)=&\frac{1}{3}\pd^2P(z)-\en P(z)-\en^2+\frac{g_2}{6}.
\]
Consequently, with the integral strategy shown in Appendix \ref{b.0}, the first two integrals
\[
\int_{\T_1}\int_{\T_2}\big(\pz\pn(\zer)\big)^2=&\t_2\intt\big(\frac{1}{6}\pd^2P(z)+\en P(z)+\en^2-\frac{g_2}{6}\big)\nn\\
=&\t_2\en\big(\pi-2\t_2\ete\big)+\t_2^2\big(\en^2-\frac{g_2}{6}\big),\label{fermioneq1}\\
\int_{\T_1}\int_{\T_2}\pn(\zer)\pz^2\pn(\zer)=&\frac{\t_2}{2}\intt\pd^2P(z)-\inter\big(\pd\pn(z_{12})\big)^2\nn\\
=&-\t_2\en\big(\pi-2\t_2\ete\big)-\t_2^2\big(\en^2-\frac{g_2}{6}\big),\label{fermioneq2}
\]
where we have utilized the integral\footnote{For the definition of $G(\e)$, please refer to Appendix \ref{b.0}. }
\[
\intt\pd^2P(z)=\limr G(\e)+\frac{i}{2}\oint_{\pd\T}\pd P(z)\dd\zb=0
\]
%and the result (\ref{pz}) shown in the boson part to derive them.

To compute the remaining  three integrations, we need to work out the following integrals first
\[
\intt\llv\pd^2P(z)\rrv^2=&\intt\pd\big(\pd P(z)\pdb^2\pb(\zb)\big)\nn\\
=&\frac{i}{2}\oint_{\pd\T}\pd P(z)\pdb^2\pb(\zb)\dd\zb+\limr G(\e)\nn\\
=&\limr G(\e)=\limr12\pi \e^{-6}.
\]
In analogy with the bosonic case, in our regularization scheme, we simply drop out the divergent part to obtain the finite answer, that is,
\[
\intt\llv\pd^2P(z)\rrv^2=0.\label{fermionpd1}
\]
Next consider the integrand $\pb(\zb)\pd^2 P(z)$
\[
\intt\pb(\zb)\pd^2P(z)=&\intt\pd\big(\pb(\zb)\pd P(z)\big)\nn\\
=&\frac{i}{2}\oint_{\pd\T}\pb(\zb)\pd P(z)\dd\zb+\limr G(\e)\nn\\
=&\frac{i}{2}\Big(\oa+\ab+\bc+\co\Big)\pb(\zb)\pd P(z)\dd\zb\nn\\
=&0.\label{fermionpd2}
\]
According to the results of \eqref{pzmode}, \eqref{fermioneq1}, \eqref{fermioneq2}, \eqref{fermionpd1} and \eqref{fermionpd2}, we can evaluate the last three integrals now, which are listed in the following
\[
&\int_{\T_1}\int_{\T_2}\left\lvert\pz\pn(\zer)\right\rvert^4\nn\\
=&\t_2\intt\llv\frac{1}{6}\pd^2P(z)+\en P(z)+\en^2-\frac{g_2}{6}\rrv^2\nn\\
=&\t_2\intt\Big(\frac{1}{36}\llv\pd^2P(z)\rrv^2+\llv\en\rrv^2\llv P(z)\rrv^2+\frac{1}{6}\big(\enbar\pb(\zb)\pd^2P(z)+\en P(z)\pdb^2\pb(\zb)\big)\nn\\
&+\big(\enbar^2-\frac{\bar{g}_2}{6}\big)\big(\pz\pn(z)\big)^2+\big(\en^2-\frac{g_2}{6}\big)\big(\pzb\pnb(\zb)\big)^2-\llv\en^2-\frac{g_2}{6}\rrv^2\Big)\nn\\
=&\t_2^2\left\lvert\en^2-\frac{g_2}{6}\right\rvert^2+\llv\en\rrv^2\big(4\t_2^2\llv\ete\rrv^2-2\pi\t_2(\ete+\eteb)\big)\nn\\
&+\Big(\t_2\en\big(\enbar^2-\frac{\bar{g}_2}{6}\big)(\pi-2\t_2\ete)+\t_2\enbar\big(\en^2-\frac{g_2}{6}\big)(\pi-2\t_2\eteb)\Big),
\]
\[
&\inter\llv\pn(\zer)\pz^2\pn(\zer)\rrv^2\nn\\
=&\inter\big(\frac{1}{2}\pd^2P(\zer)-(\pd\pn(\zer))^2\big)\big(\frac{1}{2}\pdb^2\pb(\zber)-(\pdb\pnb(\zber))^2\big)\\
=&\int_{\T_1}\int_{\T_2}\Big(\frac{1}{4}\llv\pd^2P(\zer)\rrv^2-\frac{1}{2}\pd^2P(\zer)\big(\pzb\pnb(\zber)\big)^2-\frac{1}{2}\pdb^2\pb(\zber)\big(\pz\pn(\zer)\big)^2\nn\\
+&\left\lvert\pz\pn(\zer)\right\rvert^4\Big)=\inter\llv\pd\pn(\zer)\rrv^4,
\]
\[
\inter\big(\pzb\pnb(\zber)\big)^2\pn(\zer)\pd^2\pn(\zer)=&\inter\big(\pzb\pnb(\zber)\big)^2\big(\frac{1}{2}\pd^2P(z)-(\pd\pn(\zer))^2\big)\nn\\
=&-\int_{\T_1}\int_{\T_2}\llv(\pz\pn(\zer)\rrv^4.
\]
\section{Derivation of the counterterms}\label{sec-ct}
In this Appendix, we present the derivations of the counterterms that appear in  free bosons \eqref{FBct}, free Dirac fermions \eqref{FDFct}, and free Majorana fermions \eqref{FMFct}.
It's clear that in our cases all the counterterms are proportional to $\l^2$, thus from the expression of the second-order correction of the partition function,
\[
\Z^{(2)}=&\Z^{(0)}\left(\int_{\text{T}_1^2}\int_{\text{T}_2^2}\la\L^{(1)}(x_1)\L^{(1)}(x_2)\ra-\int_{\text{T}^2}\la\L^{(2)}\ra-\frac{2}{\l^2}\int_{\text{T}^2}\la\L_{ct}\ra\right),
\label{17173}
\]
we know that $\int_{\text{T}^2}\la\L_{ct}\ra$ only need to cancel the divergent parts in $\frac{\l^2}{2}\int_{\text{T}_1^2}\int_{\text{T}_2^2}\la\L^{(1)}(x_1)\L^{(1)}(x_2)\ra$ since $\int_{\text{T}^2}\la\L^{(2)}\ra$ converges.

For free boson, we first rewrite the integrand \eqref{tttt} as
\[
\la T\bar{T}^{(0)}(z_1,\zb_1)T\bar{T}^{(0)}(z_2,\zb_2)\ra=\Big(\lvert B\lvert^2 +2A^2\Big)\lvert P(z_{12})\lvert^2+\frac{1}{4}\lvert P(z_{12})\lvert^4+...,
\]
where "..." stands for terms giving finite integral results. As shown in \eqref{b.1}, under the disk regularization \eqref{b.0}, we have\footnote{Here $\e$ represents the radius of the infinitesimal disk regulator}
\[
\Big(\lvert B\lvert^2+2A^2\Big)\int_{\text{T}_1^2}\int_{\text{T}_2^2} \lvert P(z_{12})\lvert^2=\pi\t_2\Big(\lvert B\lvert^2+2A^2\Big) \e^{-2}+\text{convergent part},
\]
\[
\frac{1}{4}\int_{\text{T}_1^2}\int_{\text{T}_2^2} \lvert P(z_{12})\lvert^4=\frac{\pi\t_2}{12}\e^{-6}+\text{convergent part},
\]
then the divergent part in $\frac{\l^2}{2}\int_{\text{T}_1^2}\int_{\text{T}_2^2}\la\L_{\text{FB}}^{(1)}(x_1)\L_{\text{FB}}^{(1)}(x_2)\ra$ is given by
\[
&\frac{\t_2\l^2}{2\pi^3}\Big(\lvert B\lvert^2+2A^2\Big) \e^{-2}+\frac{\t_2\l^2}{24\pi^3}\e^{-6}\nn\\
=&\frac{8\t_2\l^2g^2}{\pi\e^2}\la(\pd\phi)^2(\pdb\phi)^2\ra+\frac{\t_2\l^2}{24\pi^3}\e^{-6}\nn\\
=&\int_{\text{T}^2}\la\L_{\text{FB},\text{ct}}\ra=\t_2\la\L_{\text{FB},\text{ct}}\ra.\label{app,fbct}
\]
Finally, to implement the minimal subtraction, from  \eqref{app,fbct} the following choice of counterterm is the simplest one
\[
\L_{\text{FB},\text{ct}}=\l^2\cdot\left\{\frac{8g^2}{\pi\e^2}\big(\pd\phi\pdb\phi\big)^2+\frac{1}{24\pi^3\e^6}\right\},
\]

In the following, we would like to determine the counterterms of the deformed free Dirac fermion. Though the integrals given in Appendix \eqref{b.2}, we can find the divergent terms which we have omitted in the previous text \eqref{diracTTTT},
\[
\int_{\T_1}\int_{\T_2}\la\L_{\text{DF}}^{(1)}(z_1,\zb_1)\L_{\text{DF}}^{(1)}(z_2,\zb_2)\ra=&\frac{\t_2}{12\pi^3\e^6}+\frac{\t_2\lvert e_{\n-1}\lvert^2}{\pi^3\e^2}+\text{convergent part}\nn\\
=&\frac{\t_2}{12\pi^3\e^{6}}+\frac{16\t_2g^2}{\pi\e^2}\la\pd\psi^*\psi\pdb\bar{\psi}^*\bar{\psi}\ra+\text{convergent part}.
\]
According to \eqref{17173}, similar to the bosonic case, the simplest choice of Lagrangian counterterm for free Dirac fermions is
\[
\L_{\text{DF},\text{ct}}=\l^2\cdot\left\{\frac{8g^2}{\pi\e^2}\pd\psi^*\psi\pdb\bar{\psi}^*\bar{\psi}+\frac{1}{24\pi^3\e^6}\right\}.
\]

Finally, we look at the free Majorana fermion. Similar to the case of Dirac fermion,
\[
\frac{\l}{2}\int_{\T_1}\int_{\T_2}\la\L_{\text{MF}}^{(1)}(z_1,\zb_1)\L_{\text{MF}}^{(1)}(z_2,\zb_2)=&\frac{\l^2\t_2}{8\pi^3\e^2}\lvert e_{\n-1}\lvert^2+\frac{\l^2\t_2}{96\pi^3\e^6}+\text{convergent part}\nn\\
=&\l^2\frac{8\t_2g^2}{\pi\e^2}\la\pd\psi\psi\pdb\bar{\psi}\bar{\psi}\ra+\frac{\l^2\t_2}{96\pi^3\e^6}+\text{convergent part},
\]
 then the simplest form of counterterm leading to minimal subtraction  is
\[
\L_{\text{MF},\text{ct}}=\l^2\cdot\left\{\frac{8g^2}{\pi\e^2}\pd\psi\psi\pdb\bar{\psi}\bar{\psi}+\frac{1}{96\pi^3\e^6}\right\}.
\]
\section{Derivation of the $T\Tbar$-flow for 2d fermions}\label{c}
In this Appendix, we reproduce the derivation of the $T\Tbar$-flow for 2d fermionic theories as shown in \cite{Bonelli:2018kik}.

The action of the un-deformed fermionic theory living in a 2d Euclidean flat spacetime is given by
\[
\L^{(0)}=\frac{g}{2}\big(\ppsib\ga\pda\ppsi-\pda\ppsib\ga\ppsi\big)+V[\Psi].
\]
One can rewrite it in a more general form, i.e., the form in curved spacetime, which is
\[
\L^{(0)}=&\frac{g}{2}\big(\ppsib\gm\nabla_\m\ppsi-\nabla_\m\ppsib\gm\ppsi\big)+V=e^\m_{~a}X^{a}_{~\m}+V\label{cfermion1},
\]
where
\[
X^a_{~\m}:=\frac{g}{2}\big(\bar{\Psi}\gamma^a\pd_\m\psi-\pd_\m\bar{\Psi}\gamma^a\Psi\big).
\]
$X^a_{~\m}$ is independent of the metric.
We then utilize the recursion relation (\ref{Ln}--\ref{Tn}) to derive the expansion of $\L^\l$. First of all, the stress tensor of the un-deformed theory is \footnote{The formula $\frac{\pd\elc}{\pd\igmn}=\frac{1}{4}\big(e_{\m c}\delta^\l_{~\n}+e_{\n c}\delta^\l_{~\m}\big)$ is used.}
\[
\LTO=&\ema\enb\big(2\frac{\pd\LO}{\pd\igmn}-\gmn\LO\big)=2\ema\enb\frac{\pd \elc }{\pd\igmn}\xcl-\dab\LO=X_{(ab)}-\dab\LO.
\]

It is useful to introduce a new notation to mark the symmetrized tensor $\hx_{ab}:=X_{(ab)}$.
Then according to \eqref{Ln}
\[
\LE=&\frac{1}{2}\big(T^{a(0)}_{~~a}\big)^2-\frac{1}{2}
T^{a(0)}_{~~b}T^{b(0)}_{~~a}=\frac{1}{2}\Big(\TR[\hx]^2-\TR[\hx^2]+2V\TR[\hx]+2V^2\Big),
\]
from which we can derive $\LTE$, the resulting expression is
\[
\LTE=&2\ema\enb\frac{\pd\LE}{\pd\igmn}-\dab\LE=\ema\enb\Big(\frac{\pd\TR[\hx]^2}{\pd\igmn}-\frac{\pd\TR[\hx^2]}{\pd\igmn}+2V\frac{\pd\TR[\hx]}{\pd g_{\m\n}}\Big)-\dab\LE,
\]
where
\[
\frac{\pd\TR[\hx]^2}{\pd\igmn}=&2\TR[\hx]\frac{\pd\elc}{\pd\igmn}\xcl=\TR[\hx]\hx_{\m\n},\\
\frac{\pd\TR[\hx^2]}{\pd\igmn}=&\hx^a_{~b}\frac{\pd(e^{\l b} X_{a\l}+\ela\xbl)}{\pd\igmn}=\big(\hx\cdot X\big)_{(\m\n)}.
\]
Hence
\[
\LTE=&\big(\TR[\hx]+V\big)\hxab-\big(\hx\cdot X\big)_{(ab)}-\dab\LE.
\]
We continue to evaluate $\LR$
\[
\LR=&T^{a(0)}_{~~a}T^{b(1)}_{~~b}-T^{a(0)}_{~~b}T^{b(1)}_{~~a}=\TR[\hx^3]-\frac{3}{2}\TR[\hx]\TR[\hx^2]+\frac{1}{2}\TR[\hx]^3+V(\TR[\hx]^2-\TR[\hx^2])\label{l2ab},
\]
from which we finally obtain $\LTR$ as follows
\[
\LTR=&2\ema\enb\frac{\pd}{\pd\igmn}\big(\TR[\hx^3]-\frac{3}{2}\TR[\hx]\TR[\hx^2]+\frac{1}{2}\TR[\hx]^3+V(\TR[\hx]^2-\TR[\hx^2])\big)-\dab\LR,
\]
where
\[
\frac{\pd\TR[\hx^3]}{\pd\igmn}=&3\frac{\pd\hxab}{\pd\igmn}\hx_{bc}\hx_{ca}=\frac{3}{2}\big(\hx^2\cdot X\big)_{(\m\n)},\\
\frac{\pd\TR[\hx]^3}{\pd\igmn}=&3\TR[\hx]^2\frac{\pd\TR[\hx]}{\pd\igmn}=\frac{3}{2}\TR[\hx]^2\hx_{\m\n}.
\]
Therefore $\LTR$ is
\[
\LTR=&3\big(\hx^2\cdot X\big)_{(ab)}-(3\TR[\hx]+2V)\big(\hx\cdot X\big)_{(ab)}\nn\\
+&\big(\frac{3}{2}\TR[\hx]^2-\frac{3}{2}\TR[\hx^2]+2V\TR[\hx]\big)\hxab-\dab\LR\label{t2ab}.
\]
According to the nature of Grassmann variables, one actually could find two identities to reduce \eqref{l2ab} and \eqref{t2ab}, that is
\[
\TR[\hx^3]-\frac{3}{2}\TR[\hx]\TR[\hx^2]+\frac{1}{2}\TR[\hx]^3&\equiv0,\\
3\big(\hx^2\cdot X\big)_{(ab)}-3\TR[\hx]\big(\hx\cdot X\big)_{(ab)}+\frac{3}{2}\big(\TR[\hx]^2-\TR[\hx^2]\big)\hx_{ab}&\equiv\textbf{0}_{ab},
\]
where $\textbf{0}$ is the $2\times2$ null matrix. We present  all reduced results as follows
\[
\LO=&\TR[\hx]+V,\label{e67}\\
\LE=&\frac{1}{2}\TR[\hx]^2-\frac{1}{2}\TR[\hx^2]+V\TR[\hx]+V^2,\\
\LR=&V\Big(\TR[\hx]^2-\TR[\hx^2]\Big),\\
\LTO=&\hxab-\dab\LO,\\
\LTE=&(\TR[\hx]+V)\hxab-\big(\hx\cdot X\big)_{(ab)}-\dab\LE,\\
\LTR=&2V\TR[\hx]\hxab-2V\big(\hx\cdot X\big)_{(ab)}-\dab\LR\label{e72},
\]
where $\hxab$ is
\[
\hxab=\frac{g}{2}\big(\ppsib\gamma_{(a}\pd_{b)}\ppsi-\pd_{(a}\ppsib\gamma_{b)}\ppsi\big).
\]
Although we can continue to calculate the higher-order corrections, as mentioned in \cite{Bonelli:2018kik}, for the free massive fermions (i.e., $V[\Psi]=m\bar{\Psi}\Psi$), the $T\Tbar$-flow of $\L^\l$ terminates at the second-order.

The explicit forms of \eqref{e67}-- \eqref{e72}, for  massive Dirac fermions,  in complex coordinates are
\[
T^{(0)}_{zz}=&\frac{g}{2}\pseb\dpz\pse,\quad T^{(0)}_{z\zb}=-\frac{g}{4}\big(\pseb\dpzb\pse+\psrb\dpz\psr\big)-\frac{m}{2}\big(\pseb\psr+\psrb\pse\big),\quad T^{(0)}_{\zb\zb}=\frac{g}{2}\psrb\dpzb\psr,\\
T^{(1)}_{zz}=&\frac{g^2}{4}\Big(\pseb\pse\big(\pzb\pseb\pz\pse+\pz\pseb\pzb\pse\big)-(\pseb\dpz\pse)\cdot(\psrb\dpz\psr)\Big)-\frac{gm}{2}\pseb\pse(\psrb\pz\pse-\pz\pseb\psr),\\
T^{(1)}_{z\zb}=&\frac{gm}{4}\Big(\pse\psr\big(\pseb\pzb\pseb-\psrb\pz\psrb\big)-\pseb\psrb\big(\pse\pzb\pse-\psr\pz\psr\big)\Big)+m^2\pseb\pse\psrb\psr,\\
T^{(1)}_{\zb\zb}=&\frac{g^2}{4}\Big(\psrb\psr\big(\pz\psrb\pzb\psr+\pzb\psrb\pz\psr\big)-(\psrb\dpzb\psr)\cdot(\pseb\dpzb\pse)\Big)-\frac{gm}{2}\psrb\psr(\pseb\pzb\psr-\pzb\psrb\pse),\\
T^{(2)}_{zz}=&\frac{g^2m}{2}\pseb\pse\psrb\psr\big(\pz\pseb\pz\psr+\pz\psrb\pz\pse\big),\quad T^{(2)}_{z\zb}=0,\quad T^{(2)}_{\zb\zb}=\frac{g^2m}{2}\psrb\psr\pseb\pse\big(\pzb\psrb\pzb\pse+\pzb\pseb\pzb\psr\big).
\]
\[
\mathcal{L}^{(0)}=&g\big(\pseb\dpzb\pse+\psrb\dpz\psr\big)+m\big(\pseb\psr+\psrb\pse\big),\\
\mathcal{L}^{(1)}=&\frac{g^2}{2}\Big((\pseb\dpzb\pse)(\psrb\dpz\psr)+\big(\pseb\pse\pzb\pseb\pzb\pse+\psrb\psr\pz\psrb\pz\psr\big)\Big)-g^2(\pseb\dpz\pse)(\psrb\dpzb\psr)\nn\\
&-gm\Big(\pse\psr\big(\pseb\pzb\pseb-\psrb\pz\psrb\big)-\pseb\psrb\big(\pse\pzb\pse-\psr\pz\psr\big)\Big)-2m^2\pseb\pse\psrb\psr,\\
\mathcal{L}^{(2)}=&g^2m\pseb\pse\psrb\psr\Big(2\pz\pseb\pzb\psr+2\pzb\psrb\pz\pse-\pz\psrb\pzb\pse-\pzb\pseb\pz\psr\Big).
\]
Let $m=0$, the above results degenerate to the results in Section \ref{secDF}.

\bibliographystyle{JHEP}
\bibliography{BibforTTbarTorus}{}
\end{document}